\documentstyle[12pt,a4]{article}
\input epsfig.sty
\input axodraw.sty
\newcommand{\zp}[3]{Z.\ Phys.\ {\bf C#1} (19#2) #3}
\newcommand{\pl}[3]{Phys.\ Lett.\ {\bf B#1} (19#2) #3}
\newcommand{\np}[3]{Nucl.\ Phys.\ {\bf B#1} (19#2) #3}
\newcommand{\prd}[3]{Phys.\ Rev.\ {\bf D#1} (19#2) #3}

\newcommand{\eps}{\epsilon}

\def\simgt{\rlap{\lower 3.5 pt \hbox{$\mathchar \sim$}} \raise 1pt \hbox {$>$}}
\def\simlt{\rlap{\lower 3.5 pt \hbox{$\mathchar \sim$}} \raise 1pt \hbox {$<$}}

\newcommand{\beq}{\begin{equation}}
\newcommand{\eeq}{\end{equation}}
\newcommand{\bea}{\begin{eqnarray}}
\newcommand{\eea}{\end{eqnarray}}

\newcommand{\go}{\rightarrow}
\newcommand{\Nt}{\tilde{N}}
\newcommand{\lnNt}{\ln\tilde{N}}
\newcommand{\tb}{\mbox{tg$\beta$}}
\newcommand{\lsim}{\raisebox{-0.13cm}{~\shortstack{$<$ \\[-0.07cm] $\sim$}}~}
\newcommand{\gsim}{\raisebox{-0.13cm}{~\shortstack{$>$ \\[-0.07cm] $\sim$}}~}
\def\ep{\epsilon}
\def\msb{\overline{\rm MS}}

\catcode`\@=11
\def\section{\@startsection{section}{1}{\z@}{3.5ex plus 1ex minus .2ex}
{2.3ex plus .2ex}{\large\bf}}
\def\thesection{\arabic{section}.}

\def\appendix{\setcounter{section}{0}
 \def\thesection{Appendix \Alph{section}:}
 \def\theequation{\Alph{section}.\arabic{equation}}}

\def\@citex[#1]#2{\if@filesw\immediate\write\@auxout{\string\citation{#2}}\fi
  \def\@citea{}\@cite{\@for\@citeb:=#2\do
    {\@citea\def\@citea{,\penalty\@m}\@ifundefined
       {b@\@citeb}{{\bf ?}\@warning
       {Citation `\@citeb' on page \thepage \space undefined}}%
\hbox{\csname b@\@citeb\endcsname}}}{#1}}

\def\citer{\@ifnextchar [{\@tempswatrue\@citexr}{\@tempswafalse\@citexr[]}}
\def\@citexr[#1]#2{\if@filesw\immediate\write\@auxout{\string\citation{#2}}\fi
  \def\@citea{}\@cite{\@for\@citeb:=#2\do
    {\@citea\def\@citea{--\penalty\@m}\@ifundefined
       {b@\@citeb}{{\bf ?}\@warning
       {Citation `\@citeb' on page \thepage \space undefined}}%
\hbox{\csname b@\@citeb\endcsname}}}{#1}}
\relax
\begin{document}
\thispagestyle{empty}
\begin{flushright}
CERN-TH/96-231\\
DESY 96-170\\
hep-ph/9611272
\end{flushright}
\vskip 1.5cm
\begin{center}{\bf\Large\sc Soft Gluon Radiation in Higgs Boson}
\vglue .3cm
{\bf\Large\sc Production at the LHC}
\vglue 1.2cm
\begin{sc}
Michael Kr\"{a}mer$^{a,}$\footnote{Present address: 
Rutherford Appleton Laboratory, Chilton, Didcot, OX11 0QX, England}, 
Eric Laenen$^{b}$ and Michael Spira$^{b}$\\
\vglue 0.5cm
\end{sc}
$^{a}${\it Deutsches Elektronen-Synchrotron DESY, D-22603 Hamburg, FRG}
\vglue 0.05cm
$^{b}${\it CERN TH-Division, CH-1211 Geneve 23, Switzerland}
\end{center}
\vglue 1.2cm

\begin{abstract}
\noindent
We examine the contributions of soft gluons to the Higgs
production cross section at the LHC in the Standard Model and its
minimal supersymmetric extension. The soft gluon radiation
effects of this reaction share many features with the Drell-Yan
process, but arise at lowest order from a purely gluonic initial
state. We provide an extension of the conventional soft gluon
resummation formalism to include a new class of contributions which we
argue to be universal, and resum these and the usual Sudakov effects
to all orders. The effect of these new terms is striking: only if they
are included, does the expansion of the resummed cross section to
next-to-leading order reproduce the exact result to within a few
percent for the full range of Higgs boson masses.  We use our resummed
cross section to derive next-to-next-to-leading order results, and
their scale dependence.  Moreover, we demonstrate the importance of
including the novel contributions in the resummed Drell-Yan process.
\end{abstract}

\vfill
\begin{flushleft}
CERN-TH/96-231\\
DESY 96-170 \\
hep-ph/9611272 \\
November 1996
\end{flushleft}

\newpage

\setcounter{page}{1}

\noindent

\section{Introduction}

The search for Higgs particles \cite{S:higgs} is one of the most
important endeavors for future high energy $e^+e^-$ and hadron
collider experiments.  The Higgs boson is the only particle of the
Standard Model (SM) which has not been discovered so far. The direct
search at the LEP1 experiments via the process $e^+e^- \to Z^* H$
yields a lower bound on the Higgs mass of 65.2 GeV \cite{S:lep1}.
Theoretical consistency restricts the Higgs mass to be smaller than
$\sim 700$ GeV \cite{S:lattice}.  The dominant Higgs production
mechanism at the LHC, a $pp$ collider with a c.m.~energy of 14 TeV, is
the gluon fusion process $gg \to H$ which is mediated by a heavy quark
triangle loop at lowest order \cite{S:glufus}.  As an important step
to increase the theoretical precision the two-loop QCD corrections
have been calculated, resulting in a significant increase of the
predicted total cross section by about 50 -- 100\%
\cite{SDGZ,S:limit}. The dependence on the unphysical renormalization
and factorization scales decreased considerably by including these
next-to-leading-order (NLO) corrections, resulting in an estimate of
about 15\% for the remaining scale sensitivity \cite{SDGZ}.  It is
important to note, and we will demonstrate, that the NLO corrections
are dominated by soft gluon radiation effects.

The minimal supersymmetric extension of the Standard Model (MSSM) is
among the most attractive extensions of the SM.  It requires the
introduction of two Higgs doublets leading to the existence of five
scalar Higgs particles, two scalar CP-even $h,H$, one pseudoscalar
CP-odd $A$ and two charged bosons $H^\pm$. This Higgs sector can be
described by fixing two parameters, which are usually chosen to be
\tb, the ratio of the two vacuum expectation values, and the
pseudoscalar Higgs mass $M_A$. Including higher order corrections to
the Higgs masses and couplings up to the two-loop level, the mass of
the lightest scalar Higgs particle $h$ is restricted to be smaller
than $\sim 130$ GeV \cite{S:hbound}. The direct search at LEP1 sets
lower bounds of about 45 GeV on the masses of the MSSM Higgs bosons
\cite{S:lep1}.  The dominant neutral Higgs production mechanisms at
the LHC are the gluon fusion processes $gg \to h,H,A$ and the
associated production with a $b\bar b$ pair $gg,q\bar q \to b\bar b
h,b\bar b H,b\bar b A$ which becomes important only for large
\tb~\cite{S:phibb}. The coupling of the neutral Higgs particles to
gluons is again mediated by top and bottom loops, with the latter
providing the dominant contribution for large \tb, and squark loops,
if their masses are smaller than about 400 GeV \cite{S:squark}.  (In
this paper we shall assume the squark masses to be 1 TeV, so that
squark loops can safely be neglected.)  The two-loop (NLO) QCD
corrections to the gluon fusion mechanism have also been calculated
\cite{SDGZ,S:squark} and conclusions completely analogous to the SM
case emerge.  Soft gluon radiation effects again provide the dominant
contribution to these corrections, for small \tb. For large \tb~bottom
mass effects will be of similar size due to the dominance of the
bottom quark loops and we will not consider this regime in this paper.

In previous analyses the resummation of soft gluon radiation in the
transverse momentum distribution of the Higgs bosons has been
performed \cite{kauff}, which are significant at small $p_T$. We
consider universal soft gluon effects on the total production cross
section and demonstrate that these dominate the NLO corrections both
in the SM and the MSSM for small \tb.  The study of these effects in
higher orders, and their resummation to all orders is our purpose in
this paper.

Although our main focus is Higgs production, we will consider soft
gluon effects in the Drell-Yan process for comparison.  Higgs
production shares many features with this reaction, apart from the
species of leading initial state partons, e.g. it also proceeds at
lowest order via a color singlet hard scattering process, and is a
$2\go 1$ process at lowest order.  The Drell-Yan process has been
studied by performing exact perturbative QCD calculations up to
next-to-next-to-leading order (NNLO) in Ref.~\cite{DYexact} and in the
context of soft gluon resummation in \cite{ResDY}. In this paper we
present an extension of the conventional soft gluon resummation
formalism, in which we use the Drell-Yan reaction to gauge its quality
and importance.  We then apply the extension to Higgs production to
derive the first estimates of NNLO effects. These estimates are
important in view of the size of the NLO corrections.

The paper is organized as follows: In section 2 we describe the
construction of the resummed cross section and the extension of the
soft gluon resummation formalism.  In section 3 we present NLO and
NNLO results from the expansion of the resummed cross sections for
Higgs production and the Drell-Yan process.  We conclude and present
an outlook in section 4.

\section{The Resummed Exponent}

In this section we derive the resummed partonic cross section for
Higgs boson production via gluon fusion.  In order to set the stage,
we must first discuss some preliminary approximations to the exact NLO
calculation in Ref.~\cite{SDGZ}.

\begin{figure}[htb]
\begin{center}
\begin{picture}(150,90)(0,0)

\Gluon(10,20)(50,20){-3}{4}
\Gluon(10,80)(50,80){3}{4}
\ArrowLine(50,20)(50,80)
\ArrowLine(50,80)(90,50)
\ArrowLine(90,50)(50,20)
\DashLine(90,50)(130,50){2}

\put(0,18){g}
\put(0,78){g}
\put(30,46){t,b}
\put(135,46){H}

\end{picture}  \\
\caption{\label{fg:dia} \it Higgs boson production via gluon fusion mediated
by top- and bottom quark loops}
\end{center}
\end{figure}
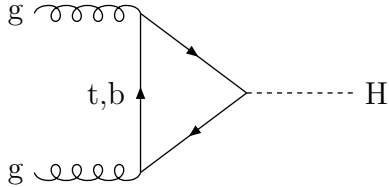

In the Standard Model, the leading order (LO) process consists of
gluon fusion into a Higgs boson via a heavy quark triangle loop, see
Fig.~\ref{fg:dia}.
Because the Higgs coupling to fermions is proportional to the fermion
mass, the top quark strongly dominates this coupling, constituting
about 90\% of the total coupling. Our {\it first} approximation is to
neglect the quark-antiquark and quark-gluon channels as these
contribute at next-to-leading order (NLO) to the full cross section
with less than 10\% \cite{SDGZ}.

The exact NLO calculation in \cite{SDGZ} was performed for the general
massive case, i.e. all masses were taken into account explicitly.
However, the following very useful approximation was identified,
involving the heavy top mass limit of the calculation. Let us define
the (NLO) K-factor\footnote{It should be noted that we use here NLO
  parton densities and strong coupling $\alpha_s$ in the NLO cross
  sections and LO quantities in the LO cross sections for this
  ``hadronic'' K-factor. This leads to K-factors smaller than two in
  contrast with the ``partonic'' K-factor, which is defined with NLO
  parton densities and strong coupling also in the LO cross section.}
by
\beq
K^{t+b}_{NLO}(\tau_t,\tau_b) \equiv \frac{\sigma_{NLO}(\tau_t,\tau_b)}
{\sigma_{LO}(\tau_t,\tau_b)}
\label{eq:kfac}
\eeq
where $\sigma_{LO/NLO}(\tau_t,\tau_b)$ denotes the hadronic
gluon-fusion cross section for the general massive case, calculated
exactly in LO/NLO, and the scaling variables are defined by $\tau_Q =
4m_Q^2/M_H^2~(Q=t,b)$.

\begin{figure}[htbp]

\vspace*{-1.2cm}
\hspace*{0.0cm}
\epsfig{file=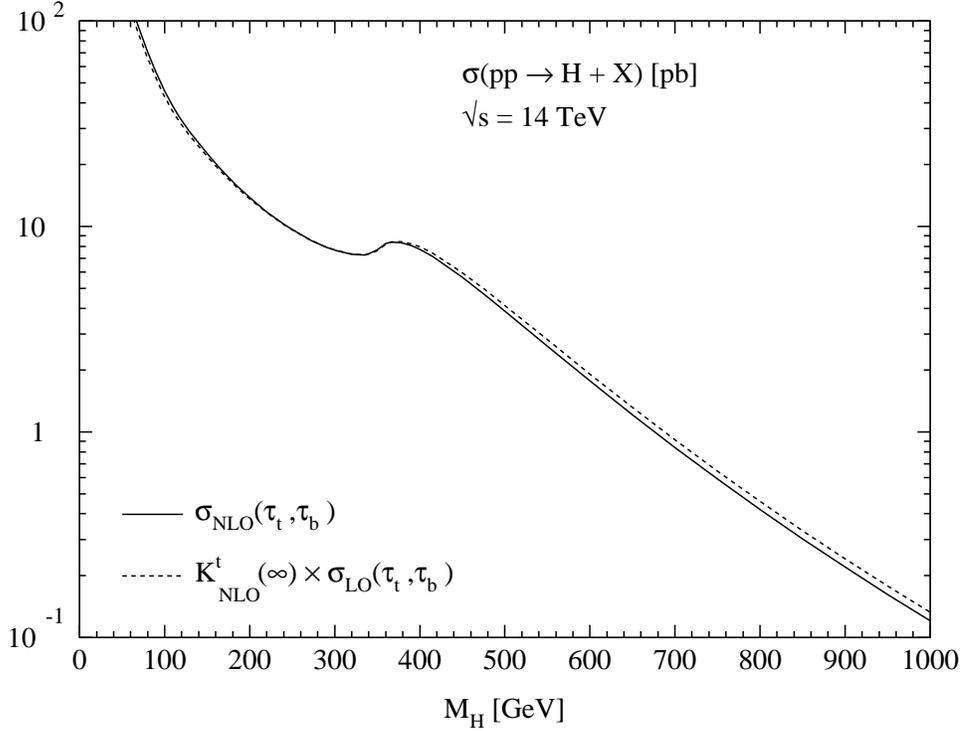,bbllx=0pt,bblly=70pt,bburx=575pt,bbury=800pt,%
        width=14.5cm,angle=-90}
\vspace*{-1.2cm}

\caption[]{\it Exact and approximate results in the heavy top quark limit
  for the total SM Higgs production cross sections as a function of
  the Higgs mass $M_H$. The top mass has been chosen as $m_t = 175$
  GeV and the bottom mass as $m_b = 5$ GeV. CTEQ4M parton densities
  \cite{CTEQ4} with NLO strong coupling [$\alpha_s(M_Z^2) = 0.116$]
  have been used.}
\label{fg:App}
\end{figure}
\begin{figure}[htbp]

\vspace*{-1.2cm}
\hspace*{0.0cm}
\epsfig{file=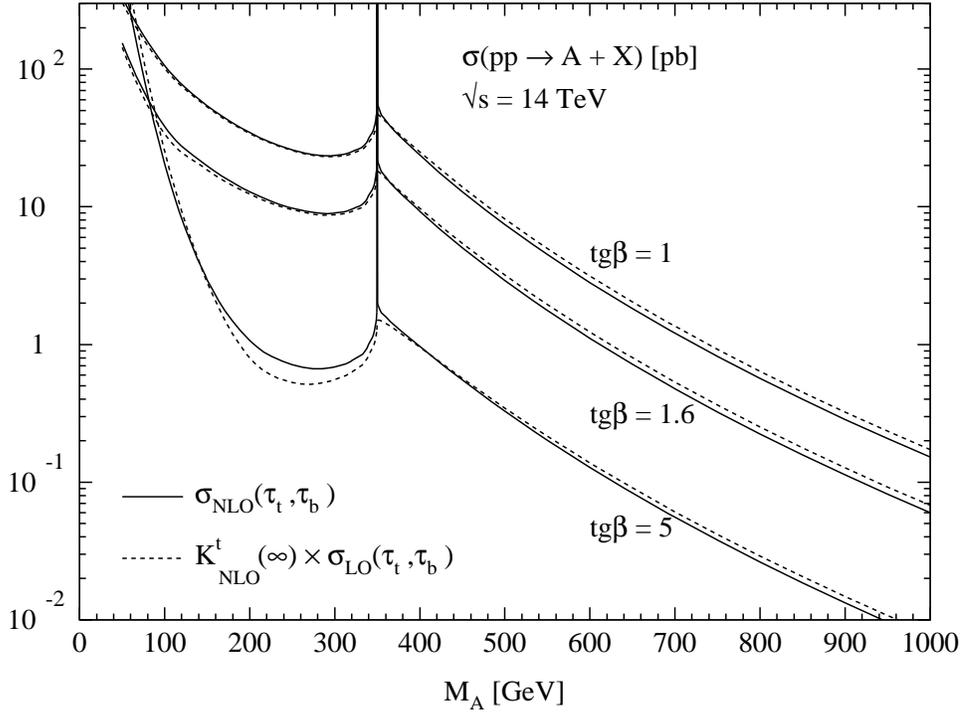,bbllx=0pt,bblly=70pt,bburx=575pt,bbury=800pt,%
        width=14.5cm,angle=-90}
\vspace*{-1.2cm}

\caption[]{\it As in Fig.\ref{fg:App}, but now for the MSSM pseudoscalar
  Higgs particle for three values of \tb.}
\label{fg:AApp}
\end{figure}
In Figs.~\ref{fg:App} and \ref{fg:AApp} we compare
$\sigma_{NLO}(\tau_t, \tau_b)$ with the approximation
\beq
K^t_{NLO}(\infty)\times  \sigma_{LO}(\tau_t,\tau_b)
\label{imassapp}
\eeq
for scalar and pseudoscalar Higgs boson production, where the K-factor
$K^t_{NLO}(\infty)$ takes into account the top quark contribution to
the relative QCD corrections only, in the limit of a heavy top quark.
We observe that the approximation (\ref{imassapp}) is accurate to
within 10\% for the full Higgs mass range $M_H \gsim 65$ GeV of the SM
Higgs boson as well as the pseudoscalar Higgs particle of the MSSM for
small \tb~[At the $t\bar t$ threshold $M_A=2m_t$ the pseudoscalar
cross section develops a Coulomb singularity so that perturbation
theory is not valid in a small margin around this value for the
pseudoscalar mass $M_A$ \cite{SDGZ}.].  The same accuracy of the
approximation emerges for the two scalar Higgs particles of the MSSM
for small \tb.  Our {\it second} approximation then consists of
assuming that $K^t(\infty)\times \sigma_{LO}(\tau_t,\tau_b)$ is a
valid approximation to $\sigma(\tau_t,\tau_b)$ for all orders, i.e.~we
will assume that the higher-order K-factor, when computed in the
infinite top mass limit and combined with the massive LO cross
section, will give a good approximation to the higher order cross
section in the general massive case.  In fact, we will see that at NLO
the bulk of the K-factor is due to soft and collinear gluons, which do
not resolve the effective coupling.  The assumption that this persists
to higher orders is supported by the validity of the infinite top mass
approximation at NLO.

In the MSSM, the validity of these approximations depends strongly on
the parameter \tb. For $\tb \lsim 1.6$ the top quark contribution to
the cross sections amounts to more than 70\%.  The heavy top quark
limit is thus a reasonable approximation in this regime. For large
values of \tb~the bottom loop contribution becomes significant so that
the approximations are no longer valid.  Still, the infinite top mass
approximation deviates from the full NLO result, including bottom
contributions, by less than 25\% for $\tb \lsim 5$ as can be inferred
from Fig.\ref{fg:AApp}.

We are now ready for the construction of the resummed partonic cross
section, for which we will employ the methods of Ref.~\cite{CLS}.  In
order to retain similarity to the Drell-Yan case, we will denote the
Higgs mass squared with $Q^2$ throughout the text of this paper.

In the approximations outlined above, the regularized total Higgs
production cross sections may be written in $d=4-2\epsilon$ dimensions
as [$\phi = h,H,A$]
\beq
\sigma^\phi (\tau_\phi,Q^2,\mu^2) = \int_{\tau_\phi}^1 dx_1
\int_{\tau_\phi/x_1}^1 dx_2~g(x_1)~g(x_2)~\hat \sigma^\phi_{gg}
(z,Q^2,\mu^2,\epsilon)
\eeq
or, in terms of moments,
\beq
\sigma^\phi(N,Q^2,\mu^2) = \int_0^1 d\tau_\phi~\tau_\phi^{N-1}~\sigma^\phi
(\tau_\phi,Q^2,\mu^2)
= g^2(N+1)~\hat\sigma^\phi_{gg} (N,Q^2,\mu^2,\epsilon)
\eeq
where $\tau_\phi = Q^2/S$, $z = \tau_\phi/(x_1 x_2)$, $S$ is the
hadronic c.m.~energy squared, $\mu$ is the dimensional regularization
scale, and $g(x)$ is the bare gluon distribution function.  Note that
the definition is such that the dependence on the moment variable in
the parton densities is shifted by one unit compared to the dependence
of the partonic cross section. In this way we remove an overall $1/z$
factor, emphasizing the soft gluon contribution to the partonic cross
section.

In the approximations discussed in the beginning of this section, the
$d$-dimensional partonic cross sections can be cast into the form
\beq
\hat\sigma^\phi_{gg} = \sigma^\phi_0~\kappa_\phi~\rho_\phi
(z,Q^2/\mu^2,\epsilon)
\label{rhodef}
\eeq
with the coefficients
\bea
\sigma^{h,H}_0 & = & g_t^{h,H}~\frac{G_F\alpha_{s,B}^2N_C C_F}{1152\sqrt{2}\pi}
\frac{\Gamma^2(1+\epsilon)}{1-\epsilon}
\left( \frac{4\pi}{m_t^2}\right)^{2\epsilon}, \\
\sigma^A_0 & = & g_t^A~\frac{G_F \alpha_{s,B}^2 N_C C_F}{512\sqrt{2}\pi}
\frac{\Gamma^2(1+\epsilon)}{1-\epsilon}
\left( \frac{4\pi}{m_t^2}\right)^{2\epsilon},
\label{sigzero}
\eea
where $\alpha_{s,B}$ is the bare strong coupling constant (with
dimension $2\epsilon$) and $g_t^\phi (\phi = h,H,A)$ denote the
modified top Yukawa couplings normalized to the SM coupling, which are
given in \cite{SDGZ}.  The factor $\kappa_\phi$ in eq.~(\ref{rhodef})
stems from the effective coupling of the Higgs boson to gluons in the
heavy top quark limit, which can be obtained by means of low energy
theorems \cite{SDGZ,S:let}.  They are given\footnote{The last two factors
in the large bracket originate from the anomalous dimension of the gluon
operators \cite{S:alsmt}. The top mass $\overline{m}_t$ denotes the scale
invariant
$\overline{\rm MS}$ mass $\overline{m}_t = \overline{m}_t(\overline{m}_t)$.
We would like to
thank the authors of Ref.~\cite{chetyr} for pointing out two errors in our
treatment of the strong coupling in eq.(\ref{eq:kapgen}) and the anomalous mass
dimension of eq.(\ref{eq:anommass}) in an earlier version of this paper. The
numerical size of these errors is about 0.03\% and thus negligible.}
by \cite{SDGZ,S:alsmt}
\bea
\kappa_{h,H} & = & \left\{ \frac{3\pi}{
\left[\alpha^{(5)}_s(\overline{m}_t^2)\right]^2}~\frac{\beta_t[\alpha^{(6)}_s
(\overline{m}_t^2)]}{1+\gamma_m[\alpha^{(6)}_s(\overline{m}_t^2)]}
\left( \frac{\alpha_s^{(5)}(\overline{m}_t^2)}{\alpha_s^{(5)}(M_{h,H}^2)}
\right)^2
\frac{\beta [\alpha_s^{(5)}(M_{h,H}^2)]}{\beta [\alpha_s^{(5)}
(\overline{m}_t^2)]} \right\}^2 \label{eq:kapgen} \\
\kappa_A & = & 1
\eea
where $\alpha_s^{(n_f)}~(n_f = 5,6)$ is the strong coupling constant in the
$\msb$ scheme including $n_f$ flavors in the evolution; the couplings for 5 and
6 flavors are related by \cite{larin}
\beq
\alpha_s^{(5)}(\overline{m}_t^2) = \alpha_s^{(6)}(\overline{m}_t^2)\left\{ 1 +
\frac{11}{72}
\left( \frac{\alpha_s^{(6)}(\overline{m}_t^2)}{\pi} \right)^2 \right\}
\eeq
$\beta(\alpha_s)$ denotes the QCD $\beta$ function and
$\beta_t(\alpha_s)$ its top quark contribution\footnote{The factors
  $\kappa_{h,H}$ include the
  top quark contribution at {\it vanishing} momentum transfer, which
  differs from the top quark contribution to the $\msb$ 
  $\beta$ function by a finite amount at ${\cal O}(\alpha_s^4)$
  \cite{bernwetz}.}, which is given by \cite{larin,betaqcd} [$n_f=5$
is the number of light flavors]
\beq
\frac{3\pi}{\alpha_s^2}~\beta_t(\alpha_s) = 
1 + \frac{19}{4}\frac{\alpha_s}{\pi}
+ \frac{6793 - 281\, n_f}{288}\frac{\alpha_s^2}{\pi^2} 
+ {\cal O}(\alpha_s^3)\,,
\eeq
$\gamma_m(\alpha_s)$ is the anomalous mass dimension including 6 flavors, which
can be expressed as \cite{anommass}
\beq
\gamma_m(\alpha_s) = 2 \frac{\alpha_s}{\pi} + \left( \frac{101}{12}
- \frac{5}{18} [n_f+1] \right) \frac{\alpha_s^2}{\pi^2} + {\cal O}(\alpha_s^3).
\label{eq:anommass}
\eeq
Using these expansions the effective scalar couplings $\kappa_{h,H}$
are given by\footnote{This result agrees with Ref.~\cite{chetyr}.}
\bea
\kappa_{h,H} & = & 1 + \frac{11}{2} \frac{\alpha^{(5)}_s(m_t^2)}{\pi}
+ \frac{3866 - 201\, n_f}{144}\left( \frac{\alpha^{(5)}_s(m_t^2)}{\pi}\right)^2
\nonumber \\
& & + \frac{153-19\, n_f}{33-2\, n_f}
~\frac{\alpha^{(5)}_s(M_{h,H}^2) - \alpha^{(5)}_s(m_t^2)}{\pi}
+ {\cal O}(\alpha_s^3)
\label{kappa}
\eea
where $m_t$ denotes the pole mass of the top quark.
The scale dependence of the strong coupling in the lowest order cross section
eq.~(\ref{sigzero}) and the factor $\rho_\phi$ eq.~(\ref{rhodef})
includes 5 light flavors, i.e.~the top quark is decoupled. In the rest of the
paper we identify
\beq
\alpha_s (\mu^2) \equiv \alpha_s^{(5)} (\mu^2) \, .
\eeq

Note further that the expression (\ref{rhodef}) is not yet finite for
$d\go 4$; mass factorization and renormalization of the bare coupling
in the Born cross section will be carried out after resummation.  In
eq.~(\ref{rhodef}) we denote the {\it correction factors} by
$\rho_\phi(z,Q^2/\mu^2,\epsilon)$, which are defined in the infinite
mass limit {\it without} the factorizing corrections $\kappa_\phi$ to
the effective coupling.  They may be expanded as
\beq
\rho_\phi(z,Q^2/\mu^2,\alpha(\mu^2),\epsilon) = \sum_{n=0}^{\infty}~
\alpha^n(\mu^2) \rho_\phi^{(n)}(z,Q^2/\mu^2,\epsilon)
\label{sumpth}
\eeq
where we define, for the sake of convenience,
\beq
\alpha(\mu^2) \equiv \frac{\alpha_s(\mu^2)}{\pi}\, .
\eeq
Here $\alpha_s(\mu^2)$ is the renormalized strong coupling
and we have chosen the renormalization scale equal to 
$\mu$ for the moment.
From Ref.~\cite{SDGZ,S:limit}
we can derive the following expression
for the first two coefficients of the SM Higgs correction factor
\begin{eqnarray}
\rho_\phi^{(0)}(z,Q^2/\mu^2,\epsilon) & = & \delta(1-z) \\
\rho_{h,H}^{(1)}(z,Q^2/\mu^2,\epsilon) & = & 
\Big(\frac{\mu^2}{Q^2}\Big)^\ep C_A\Big\{ -\frac{z^\epsilon}{\epsilon}
\left[\frac{1+z^4+(1-z)^4}
{(1-z)^{1+2\epsilon}}\right]_+ \nonumber \\
&+&\delta(1-z)\left(\frac{11}{6\epsilon}
 + \frac{203}{36} + \frac{\pi^2}{3}\right)
-\frac{11}{6} z^\epsilon (1-z)^{3-2\epsilon} \Big\}
\label{rho1} \\
\rho_A^{(1)} & = & \rho_{h,H}^{(1)} + 2 \left(\frac{\mu^2}{Q^2}
\right)^\eps C_A \delta(1-z)
\label{rho1a}
\end{eqnarray}
Note that we have implicitly redefined the scale $\mu$ by
$\mu^2 \rightarrow \mu^2\exp [-(\ln(4\pi) - \gamma_E)]$
to eliminate  factors $(4\pi)^{\epsilon}$
and $\Gamma(1-\eps)/\Gamma(1-2\eps)$.
The plus distribution in eq.~(\ref{rho1}) is as usual defined by
\beq
\int_x^1dz g(z)[f(z)]_+ = \int_x^1dz(g(z)-g(1))f(z)-g(1)\int_0^xdzf(z)\,.
\label{plus}
\eeq

We will now construct a resummed expression for
$\rho_\phi(z,Q^2/\mu^2,\alpha,\epsilon)$ by means of the methods
described in Ref.~\cite{CLS}.  Near the elastic edge of phase space
the Higgs cross section in the infinite mass limit may be factorized
into hard, soft and jet functions, in completely analogy with the
Drell-Yan cross section. Following the arguments of Ref.~\cite{CLS}
this leads to the Sudakov evolution equation
\beq
Q^2 \frac{d}{dQ^2} \rho_\phi(z,Q^2/\mu^2,\alpha(\mu^2),\epsilon)\! = \!
   \int^1_z \frac{dz'}{ z'} W_\phi(z',Q^2/\mu^2,\alpha(\mu^2),\epsilon) 
\rho_\phi(z/z',Q^2/\mu^2,\alpha(\mu^2),\epsilon)\, .
\label{sudevz}
\eeq
In order to solve eq.~(\ref{sudevz}) we must impose a boundary condition.
We will shortly argue \cite{CLS} that we may use the boundary condition
$\tilde \rho_\phi(N,Q^2/\mu^2=0,\alpha(\mu^2),\epsilon)=1$ for the moments,
or in $z$-space
\beq
\rho_\phi(z,Q^2/\mu^2=0,\alpha(\mu^2),\epsilon) = \delta(1-z). \label{bc}
\eeq
The solution to eq.~(\ref{sudevz}) is then, in Mellin space,
\beq
\tilde \rho_\phi(N,Q^2/\mu^2,\alpha(\mu^2),\ep) 
= \exp\left[ \int_0^{Q^2}
 {d \xi^2 \over \xi^2} 
{\tilde W}_\phi \biggl(N,{\xi^2\over \mu^2},\alpha(\mu^2),\ep\biggr) 
\right]\, ,
\label{sudsolveN}
\eeq
where as above $d=4-2\epsilon>4$. The formal Mellin inversion reads
\beq
W_\phi(z,Q^2/\mu^2,\alpha(\mu^2),\epsilon) = \int_{-i\infty}^{i\infty}
\frac{dN}{ 2\pi i}\, N^{-z}\;
{\tilde W}_\phi(N,Q^2/\mu^2,\alpha(\mu^2),\epsilon)\, .
\eeq
The solution (\ref{sudsolveN}) may be expressed as
\begin{eqnarray}
{\tilde \rho}_\phi(N,Q^2/\mu^2,\alpha(\mu^2),\epsilon)& = &
\exp\biggl[\int_0^1dzz^{N-1}\int_0^{\nu Q^2/\mu^2}\frac{d\lambda}{\lambda}
\Big\{\alpha(\lambda,\alpha(\mu^2),\epsilon)W_\phi^{(1)}(z,1/\nu,\epsilon)
\nonumber\\
&+&\alpha^2(\lambda,\alpha(\mu^2),\epsilon)W_\phi^{(2)}(z,1/\nu,\epsilon) +
{\cal O}(\alpha^3) \Big\}
\biggr]\ .
\label{eqnsoln}
\end{eqnarray}
with $\nu(z)$ an arbitrary function. Note that $\nu(z)$ is to be
treated as part of any plus distributions in $W_\phi$, because it
arises from the originally $z$-independent scale ratio.  In deriving
eq.~(\ref{eqnsoln}) we have used the renormalization group invariance
of the evolution kernel $W_\phi$.  Note further that we have expanded
the full evolution kernel $W_\phi(z,\xi^2/\mu^2,\alpha(\mu^2),\ep)$ in
the $d$-dimensional {\it running} coupling constant, with
$\lambda=\nu(z)\xi^2/\mu^2$.  The defining equation of the
$d$-dimensional running coupling is
\beq
\lambda^{1-\epsilon}{\partial 
[\lambda^\epsilon\alpha(\lambda,\alpha(\mu^2),\epsilon)]
\over \partial \lambda}=-b_2\alpha^2(\lambda,
\alpha(\mu^2),\epsilon)-
b_3\alpha^3(\lambda,
\alpha(\mu^2),\epsilon)\, ,
\label{dalfas}
\eeq
with the boundary condition
$\alpha(1,\alpha(\mu^2),\epsilon)=\alpha(\mu^2)$.  Here $b_2 = (11 C_A
- 2 n_f)/12$ and $b_3 = 34 C_A^2/48 - (20 C_A/3 + 4 C_F) n_f/32$.  The
solution, linearized in $b_3$, is
\beq 
\lambda^\epsilon\alpha(\lambda,\alpha(\mu^2),\epsilon)=
{\alpha(\mu^2)\over 1-\gamma(\lambda^\epsilon,\epsilon)
\alpha(\mu^2)}+{b_3\over b_2}
{\alpha^2(\mu^2)\over [1-\gamma(\lambda^\epsilon,\epsilon)
\alpha(\mu^2)]^2}f(\lambda^\epsilon,\alpha(\mu^2),\epsilon)\ ,
\label{dalfasdef}
\eeq
with
$\gamma(\lambda^\epsilon,\epsilon)\equiv {b_2\over
\epsilon}(\lambda^{-\epsilon}-1),\ 
f(\lambda^\epsilon,\alpha,\epsilon)=1-\lambda^{-\epsilon}-
\left(1+{\epsilon\over b_2\alpha}\right)
\ln[1-\gamma(\lambda^\epsilon,\epsilon)\alpha]$.
We shall now justify the boundary condition in eq.~(\ref{bc}).
To this end, note that for $\epsilon<0$ ($d>4$), the 
dimensionally-continued running coupling vanishes at zero scale,
$\alpha(0,\alpha(\mu_1^2),\epsilon)=0$, order by order in its perturbative
expansion in the coupling $\alpha(\mu_1^2)$ evaluated at any
nonzero scale $\mu_1$.  Dimensionally-continued radiative
corrections to $\rho_\phi$ therefore vanish at $Q^2=0$.

The functions $W_\phi^{(1)},\ W_\phi^{(2)}$ 
can be determined by choosing $\nu=1$ in eq.~(\ref{eqnsoln})
and expanding the Sudakov equation to second order in the 
$d$-dimensional running coupling
\bea
Q^2{\partial \over \partial Q^2}
\ln\; \tilde \rho_\phi(N,Q^2/\mu^2,\alpha(\mu^2),\epsilon)
&=& 
\alpha(Q^2/\mu^2,\alpha(\mu^2),\epsilon)\,{\tilde W}_\phi^{(1)}(N,1,\epsilon) 
 \nonumber \\
&&\hbox{\hskip -1.0 true in}+ \alpha^2(Q^2/\mu^2,\alpha(\mu^2),\epsilon)
\,{\tilde W}_\phi^{(2)}(N,1,\epsilon)\, ,
\eea
in terms of $\alpha_s(\mu^2)$, using (see eq.~(\ref{dalfasdef}))
\beq
\lambda^{\epsilon}\alpha(\lambda,\alpha(\mu^2),\epsilon)=
\alpha(\mu^2)
+\gamma(\lambda^\epsilon,\epsilon)\alpha^2(\mu^2) + {\cal O}(\alpha^3)\ .
\eeq

The resulting
one- and two-loop coefficients of the evolution kernel 
can then be derived from low order calculations of the correction factors
via
\beq
{\tilde W}_\phi^{(1)}(N,1,\epsilon)=
(Q^2/\mu^2)^{\epsilon}\, 
Q^2 \frac{\partial }{ \partial Q^2}\, 
{\tilde \rho}_\phi^{(1)}(N,Q^2/\mu^2,\epsilon)\ ,
\eeq
and 
\bea
{\tilde W}_\phi^{(2)}(N,1,\epsilon) &=& 
(Q^2/\mu^2)^{2\epsilon}\, \Bigg \{
\frac{\partial }{ \partial \ln Q^2}\, \biggl (
{\tilde \rho}_\phi^{(2)}(N,Q^2/\mu^2,\epsilon) 
-\frac{1}{ 2} \biggl [{\tilde \rho}_\phi^{(1)}(N,Q^2/\mu^2,\epsilon)
\biggr ]{^2}
\biggr ) \nonumber\\
&& \hbox{\hskip 1.0 true in}
- \gamma((Q^2/\mu^2)^\epsilon,\epsilon)\frac{\partial }{ \partial \ln Q^2}
{\tilde \rho}_\phi^{(1)}(N,Q^2/\mu^2,\epsilon) \bigg \}\, .
\eea

Of course, unlike for Drell-Yan \cite{DYexact}, the second order
corrections to Higgs production have not yet been calculated.  The
above results can easily be inverted to derive
$W^{(1,2)}_\phi(z,1,\epsilon)$, if the functions $\rho_\phi^{(1,2)}$
are known.  The full functions
$W_\phi^{(1,2)}(z,\xi^2/\mu^2,\epsilon)$ may then be constructed by
reexpanding the running coupling
$\alpha(\xi^2/\mu^2,\alpha(\mu^2),\ep)$ in $\alpha(\mu^2)$, using
eq.~(\ref{dalfasdef}) with $\lambda=\xi^2/\mu^2$,
\bea
W_\phi^{(1)}(z,\xi^2/\mu^2,\epsilon)
&=&
\bigg ({\mu^2 \over \xi^2}\bigg)^\epsilon W_\phi^{(1)}(z,1,\epsilon)\, ,
\label{wrescale}
\\
W_\phi^{(2)}(z,\xi^2/\mu^2,\epsilon)&=& 
\bigg ({\mu^2 \over \xi^2}\bigg)^{2\epsilon} W_\phi^{(2)}(z,1,\epsilon)
+\gamma((\xi^2/\mu^2)^\epsilon,\epsilon)\,
\bigg ({\mu^2 \over \xi^2}\bigg)^\epsilon W_\phi^{(1)}(z,1,\epsilon)\, .
\nonumber
\eea

In this way the one-loop coefficients $W_\phi^{(1)}$ can be
straightforwardly determined from eqs.~(\ref{rho1},\ref{rho1a}). They may
be written as
\beq
W_\phi^{(1)}(z,1,\ep)=\delta(1-z)f_\phi^{(1)}(\ep)
+z^\ep\left({g^{(1)}(z,\ep)\over (1-z)^{1+2\ep}}\right)_+
+h^{(1)}(z,\ep)\ ,
\label{wdecomp}
\eeq
where the coefficient functions $f_\phi^{(1)},\ g^{(1)},
\ h^{(1)}$
are regular functions of their arguments at $z=1$.
In the present case they are given by
\bea
f_{h,H}^{(1)}(\ep) & = & -C_A\ep\biggl(\frac{11}{6\ep} + {203\over 36}+
{\pi^2\over 3}\biggr)\ ,
\label{f1} \\
f_A^{(1)}(\ep) & = & f_{h,H}^{(1)}(\eps) - 2 C_A\ep ,
\label{f1a} \\
g^{(1)}(z,\ep) & = &
C_A\biggl(1+z^4+(1-z)^4\biggr)\, ,
\label{g1} \\
h^{(1)}(z,\ep) & = & C_A \eps \frac{11}{6} z^\eps (1-z)^{3-2\eps}.
\label{h1}
\eea
Given the limitations of the factorization theorem near the edge of
phase space \cite{St}, from which the evolution equation
(\ref{sudevz}) is derived \cite{CLS}, we might be tempted to
immediately discard all terms of order $1/N$.  However, we wish to be
careful with these terms, and we will examine them and their relevance
later in this section.

As was shown in \cite{CLS}, the term $\delta(1-z)\; f_\phi^{(1)}$ and
the plus-distributions in eq.~(\ref{wdecomp}) are separately
renormalization group invariant.  We are therefore free to choose
different functions $\nu(z)$ for these two terms in the general
resummed expression eq.~(\ref{eqnsoln}).  The natural choice for the
$\delta(1-z)\; f_\phi^{(1)}$ term is $\nu=1$. Changes in $\mu$
generate terms $b_2\ln(\mu'/\mu)$ at higher orders.  The $\lambda$
integral may then be carried out explicitly for this term in
eq.~(\ref{eqnsoln}).  For the plus-distribution term however, the
natural choice is $\nu(z)=(1-z)^2$, just as for Drell-Yan.  Then,
using $\xi=(1-z)\mu$ in eq.~(\ref{wrescale}), we absorb the factor
$(1-z)^{-2\epsilon}$ in (\ref{wdecomp}) into the boundary of the
$\lambda$ integral in eq.~(\ref{eqnsoln}), involving only the running
coupling. The term $h^{(1)}$ is of ${\cal O}(1/N^4)$ and will be
neglected.

We now turn our attention to the function $g^{(1)}(z,\ep)$ in
(\ref{g1}). We have some freedom in its treatment, as different
choices will only differ by constants or in $O(\ln N/N)$.  However, we
will argue that among all such terms the ones generated by
$g^{(1)}(z,\ep)$ are universal and can legitimately be included in the
resummed expression.  To exhibit the importance of the different
treatments of the residue function $g^{(1)}(z,\ep)$ we choose three
schemes which probe the full range of possibilities.  After rescaling
to incorporate the factor $(1-z)^{-2\ep}$ and combining the plus
distribution with the Mellin transform in eq.~(\ref{eqnsoln}), we see
that the relevant function to approximate is
$(z^{N-1}-1)g^{(1)}(z,\ep)$ [We are neglecting the coefficient
$z^\ep$, because it is contributing at ${\cal O}(1/N)$.].  The three
schemes are defined by
\bea
\mbox{scheme~} \alpha &:& \frac{1}{C_A}
(z^{N-1}-1)g^{(1)}(z,\ep)\go (z^{N-1}-1)~2
\nonumber\\
\mbox{scheme~} \beta &:& \frac{1}{C_A}
(z^{N-1}-1)g^{(1)}(z,\ep) \go (z^{N-1}-1)~2 
   - (1-z)(2z^2-4z-2z^3) \nonumber\\
\mbox{scheme~} \gamma &:& \frac{1}{C_A}
(z^{N-1}-1)g^{(1)}(z,\ep) \go (z^{N-1}-1)~2
   - (1-z)(2z^2-4z-2z^3) \nonumber \\
& & \hspace*{7.0cm} - 4 z^{N-1} (1-z)\,. 
\label{schemes}
\eea
The minimal scheme $\alpha$ involves replacing $g^{(1)}(z,\ep)$ simply
by $g^{(1)}(1,\ep)$, scheme $\beta$ includes all terms of ${\cal
  O}(1)$ in the exponent, whereas scheme $\gamma$ includes in addition
some $O(\ln^i N/N)$ terms in the exponent.  Using the one-loop
evolution kernel $W_\phi^{(1)}$ we can now construct the resummed
expressions for the Higgs production correction factor in the three
schemes. However, these expressions are still divergent for $\ep\go
0$. The divergences are cancelled by mass factorization and
renormalization for which we choose the $\msb$ scheme \cite{msbar}
throughout this paper.  For this purpose we need the resummed $\msb$
gluon distribution \cite{CSpdf}. To one loop accuracy it may be
written as \cite{CLS,CSpdf}
\bea
{\tilde \phi}^{(1)}_{\msb} \left(N,\frac{M^2}{\mu^2},\alpha(\mu^2),
\epsilon\right) 
& = & \exp\left[ -\int_0^{M^2}{d\mu'{^2}\over \mu'{^2}}\, 
\Gamma^{(1)}_{gg}(N,\alpha_s(\mu'{^2}))
\;\right ] \nonumber \\
& = &
\bar\phi^{(1)}_{\msb}\left(N,\frac{M^2}{\mu^2},\alpha(\mu^2),\eps \right)~
z_0\left(\frac{M^2}{\mu^2},\alpha(\mu^2),\eps \right)
\label{msbsplit}
\eea
where
\bea
 \bar\phi^{(1)}_{\msb}&=&\exp\left[{C_A\over 2}
\int_0^1dz\left({z^{N-1}-1\over 1-z}\right)\; (1+z^4+(1-z)^4)
 \int_0^{M^2/\mu^2}{d\lambda\over \lambda} \;\alpha(\lambda,\alpha
(\mu^2),\epsilon) \right] \nonumber\\
&\times& \exp\left[-\frac{11}{6}{C_A\over 2}\int_0^{M^2/\mu^2}
{d\lambda\over \lambda} \;\alpha(\lambda,\alpha(\mu^2),\epsilon)
\right]\ .
\label{msbdist}
\eea
and 
\beq
z_0\left(\frac{M^2}{\mu^2},\alpha(\mu^2),\eps \right)
= \exp\left[b_2\int_0^{M^2/\mu^2}
{d\lambda\over \lambda} \;\alpha(\lambda,\alpha(\mu^2),\epsilon)
\right],
\eeq
where $M$ is the mass factorization scale.  As is well-known, the
one-loop anomalous dimension $\Gamma^{(1)}_{gg}(N,\alpha_s)$ is
derived from the residue of the collinear singularity in the gluon
operator matrix element. The function $z_0$ is related to that
component of the residue which is proportional to the one loop
coefficient of the QCD $\beta$-function.  The strong coupling in the
LO cross section (\ref{sigzero}) has been left unrenormalized so far.
The renormalization of this bare coupling is now performed in the
$\msb$ scheme, via the replacement
\bea
\alpha_{s,B}  &=& \alpha_s(R^2)\,\left(R^2\right)^\eps\,
Z_\alpha\left(\frac{R^2}{\mu^2}, \alpha_s(\mu^2), \eps \right)
\label{eq:zfac} \\
&=& \alpha_s(R^2)\,\left(R^2\right)^\eps\,
\exp\left[\int_0^{R^2/\mu^2}
{d\lambda\over \lambda}\left\{b_2\,\alpha(\lambda,\alpha(\mu^2),\epsilon)
+ b_3\,\alpha^2(\lambda,\alpha(\mu^2),\epsilon)+\ldots\right\}
\right] \nonumber
\eea
where we explicitly show the renormalization scale $R$.  Thus the
renormalization factorizes into several pieces according to the
perturbative expansion of the QCD $\beta$ function,
\beq
\alpha_{s,B} = \alpha_s(R^2)\,\left(R^2\right)^\eps\,
z_0\left(\frac{R^2}{\mu^2}, \alpha_s(\mu^2), \eps \right)~
z_1\left(\frac{R^2}{\mu^2}, \alpha_s(\mu^2), \eps \right)\ldots
\label{bcren}
\eeq
where
\beq
z_1\left(\frac{R^2}{\mu^2},\alpha(\mu^2),\eps \right)
= \exp\left[b_3\int_0^{R^2/\mu^2}
{d\lambda\over \lambda} \;\alpha(\lambda,\alpha(\mu^2),\epsilon)
\right].
\eeq
Note that the above form of the $Z$-factor for the strong coupling
constant, which is completely factorized from the correction factor,
is very similar in form to the $\delta(1-z)$ piece of the $\msb$
density, and we may thus combine the overall coupling constant
renormalization with the mass factorization procedure. Restricting
ourselves to next-to-leading order (i.e. putting $z_1=1$) and choosing
$R=M$ one finds that simultaneous mass factorization and
renormalization of the bare coupling in eq.~(\ref{sigzero}) leads to
\beq
\rho_\phi(N,Q^2/M^2,\alpha(M^2)) = 
{\rho_\phi(N,Q^2/\mu^2,\alpha(\mu^2),\ep) \over
\Big[{\bar\phi}^{(1)}_{\msb}(N,M^2/\mu^2,\alpha(\mu^2),\ep)\Big]^2}\, .
\label{massfact}
\eeq
Note that the dependence on the dimensional regularization scale $\mu$
drops out on the l.h.s.  Because $\rho_\phi$ is infrared safe, we can
return to four dimensions, and we find
\bea
\rho_{h,H}^{\alpha}\left(N,\frac{Q^2}{M^2},\alpha(M^2)\right) \!\!\!\! &=&
\!\!\!\! \exp\biggl[-C_A\int_0^1 dz\frac{z^{N-1}-1}{1-z}
2\int_{(1-z)^2\frac{Q^2}{M^2}}^1\frac{d\lambda}{\lambda}
\alpha(\lambda,\alpha(M^2),\epsilon)\biggr] \nonumber\\
&\times & \!\!\!\! \exp\biggl\{\alpha(Q^2)C_A \biggl[\pi^2/3+203/36-11/6\,
\ln\left(\frac{Q^2}{M^2}\right) \biggr]
\nonumber \\
&& \quad\quad 
- 11/12\,\alpha^2(Q^2)C_Ab_2\ln^2\left(\frac{Q^2}{M^2}\right)
\biggr\} , \label{rhoa} \\
\rho_{h,H}^{\beta}\left(N,\frac{Q^2}{M^2},\alpha(M^2)\right) \!\!\!\! &=&
\!\!\!\! \rho_{h,H}^{\alpha}\left(N,\frac{Q^2}{M^2},\alpha(M^2)\right)
\\
&\times & \!\!\!\! \exp\biggl[-2C_A\!\int_0^1\!dz(2z-z^2+z^3)\!
\int_{(1-z)^2\frac{Q^2}{M^2}}^1\frac{d\lambda}{\lambda}
\alpha(\lambda,\alpha(M^2),\epsilon) \biggr] \,,
\nonumber \\
\rho_{h,H}^{\gamma}\left(N,\frac{Q^2}{M^2},\alpha(M^2)\right) \!\!\!\! &=&
\!\!\!\! \rho_{h,H}^{\beta}\left(N,\frac{Q^2}{M^2},\alpha(M^2)\right)
\label{rhoc}
\\
&\times & \!\!\!\! \exp\biggl[+4C_A\int_0^1dz~z^{N-1}
\int_{(1-z)^2\frac{Q^2}{M^2}}^1\frac{d\lambda}{\lambda}
\alpha(\lambda,\alpha(M^2),\epsilon) \biggr] \,,
\nonumber \\
\rho_A^{\alpha,\beta,\gamma}\left(N,\frac{Q^2}{M^2},\alpha(M^2)\right)
\!\!\!\! &=& \!\!\!\!
\rho_{h,H}^{\alpha,\beta,\gamma}\left(N,\frac{Q^2}{M^2},\alpha(M^2)\right)
\times \exp \biggl[ 2 C_A \alpha(Q^2) \biggr] \, .
\label{rhoA}
\eea
A few remarks are in order. First, the above expressions are formally
not well defined because the integration paths in the exponent
traverse a singularity, related to an infrared renormalon. Our purpose
in this paper is however not the numerical evaluation of the {\it
  resummed} cross section; we consider the resummed formulae to be a
generating functional for an approximation to the QCD perturbation
series, rather than its approximate sum.

The second remark is that one may try to incorporate (a part
of) the two loop evolution kernel. Some of the NNLO terms
in the exact calculation can be inferred using renormalization
group methods, such as used for Drell-Yan in Ref.~\cite{DYRG}.
A straightforward calculation along these lines leads to 
\beq
W_\phi^{(2)}(z,1,\eps) = z (1-z)^{-4\eps} P^{(1)}_{gg}(z) + \ldots,
\label{W2}
\eeq
where $P^{(1)}_{gg}$ is the two-loop component of the 
Altarelli-Parisi gluon-to-gluon splitting function \cite{pgg}.
Moreover, the mass factorization and renormalization 
of the overall coupling in the Born cross section can be
carried out in a similar fashion as at one loop, leading
to the modified $\msb$ distribution (for $R=M$)
\beq
\bar\phi^{(2)}_{\msb}(N,\frac{M^2}{\mu^2},\alpha(\mu^2),\eps) = 
\phi^{(2)}_{\msb}(N,\frac{M^2}{\mu^2},\alpha(\mu^2),\eps)~
z_0^{-1}(\frac{M^2}{\mu^2},\alpha(\mu^2),\eps)~
z_1^{-1}(\frac{M^2}{\mu^2},\alpha(\mu^2),\eps)\ldots
\eeq
where $\phi^{(2)}_{\msb}$ denotes the resummed $\msb$ distribution 
based on the two-loop anomalous dimension $\Gamma^{(2)}_{gg}$
\cite{pgg}. However, we stress that in this paper we restrict
ourselves to examining the consequences of resummation with 
the one-loop kernel $W_\phi^{(1)}$.\footnote{Except for a
$b_2\ln(\mu'/\mu)$ term at NNLO.}

There are several arguments supporting the inclusion of $\ln^iN/N$
terms in the resummation via the $\gamma$-scheme. In momentum space
such terms correspond either to the usual plus-distributions, which
are already included, or to terms $\ln^i(1-z)$, see the Appendix.  The
latter are usually discarded in Sudakov resummation procedures.  Their
inclusion extends the resummation formalism to incorporate subleading,
in the sense that they are down by a power of $N$ compared to the plus
distributions, divergent contributions for $z\go 1$ in addition to the
plus distributions.

Our first observation is that those $\ln^i(1-z)$ terms which are
included by using scheme $\gamma$ originate from the Altarelli-Parisi
splitting function, as we have demonstrated above. In distinction to
the plus-distributions, which have the same origin, they are not
enhanced by the infrared eikonal factor $1/(1-z)$, hence they are
essentially of a collinear nature. Their origin thus implies that they
are universal and independent of the hard process.  A second
observation is that the integral in the exponent in eq.~(\ref{rhoc}),
which generates these logarithms, is partly composed of the full
$\msb$ distribution (\ref{msbdist}), which is an {\it exact} result,
without any soft gluon approximation.  A third supporting argument can
be obtained from existing NNLO ($\msb$ scheme) calculations for the
$q\bar{q}$ channel in the Drell-Yan reaction \cite{DYexact} and the
non-singlet structure function $F_2$ \cite{DIS} in deep-inelastic
scattering.  We are able to derive the leading $\ln^i(1-z)$ terms and
those subleading terms which are related to the running coupling at
NNLO using the resummed formulae for the hard parts for these
reactions, given in \cite{ResDY,CLS} (e.g. for Drell-Yan see
eq.~(5.18) and the analogous DIS result obtained from renormalizing
eq.~(5.13) in \cite{CLS}).

Numerical consequences of including the $\ln^i(1-z)$ terms will be
presented in the next section for both Drell-Yan and Higgs boson
production at the LHC.

\section{Two- and Three Loop Results}

In this section we examine the quality and the scheme dependence of
our resummed formulae. We do this by expanding our results for the
correction factors of Higgs boson production and the analogous
expression for the $q\bar{q}$ channel in Drell-Yan (eq.~(5.18) in
\cite{CLS}), in various schemes ($\alpha,\beta,\gamma$) to NLO and
NNLO in order to compare with the exact results of Refs.~\cite{SDGZ}
and \cite{DYexact}. All comparisons will be made for the LHC.  We will
find that for both reactions scheme $\gamma$ reproduces the exact
results remarkably well, much better than the more conventional
schemes $\alpha$ and $\beta$ at low Higgs/$\gamma^*$ masses. We
further derive the scale-dependent logarithms in the NNLO expansion
from our resummed expressions. Here also the agreement is reasonable.

From the NNLO expansion of our resummed formulae for Higgs production,
we can obtain the first indication on the importance of the NNLO
contributions.  As mentioned earlier, these estimates are important
because the NLO K-factor is sizable.  A full numerical study of the
{\it resummed} cross sections, which requires a regularization of the
renormalon singularity \cite{renormalon}, will be published elsewhere
\cite{KLS2}.

\begin{figure}[htbp]

\vspace*{-1.1cm}
\hspace*{0.0cm}
\epsfig{file=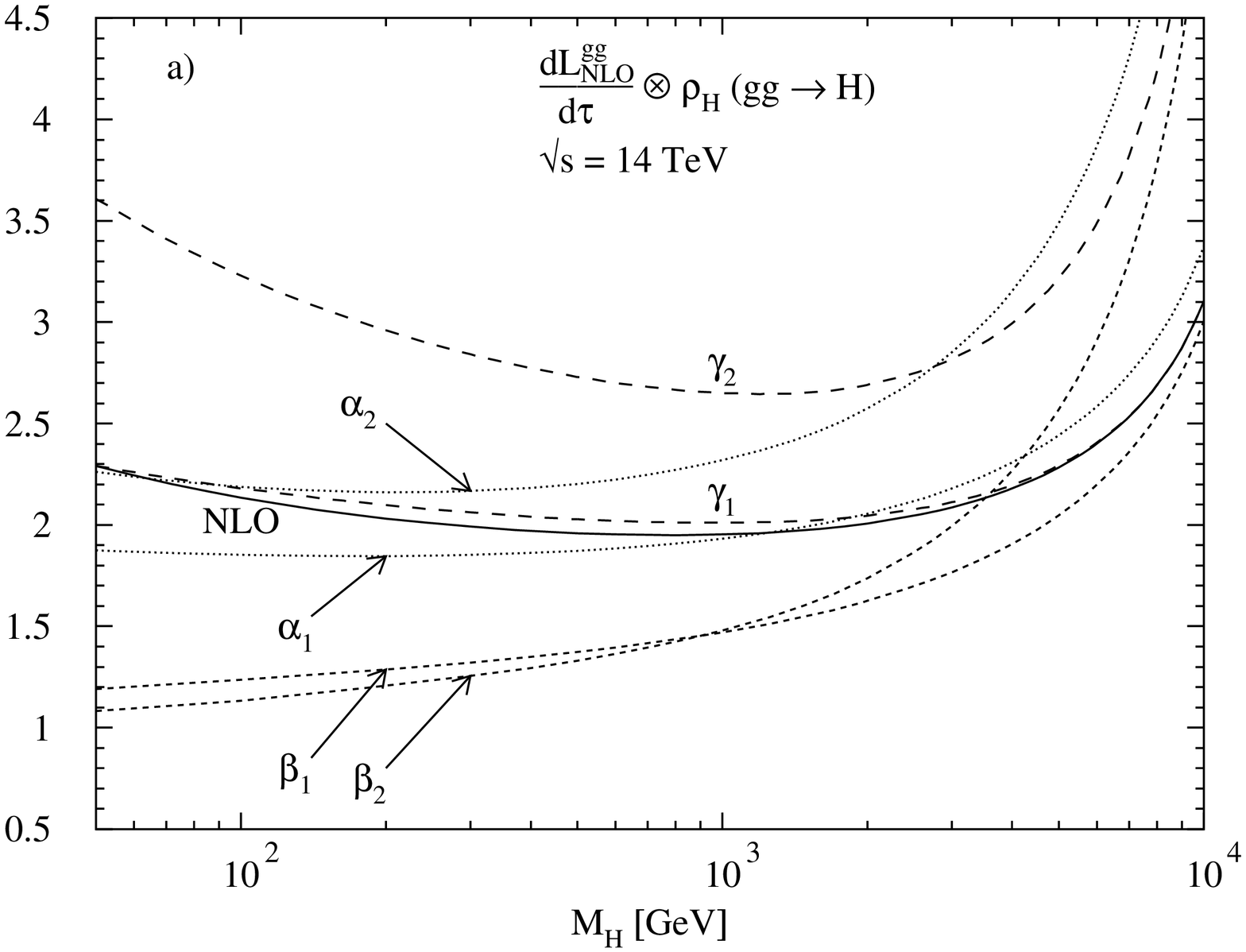,bbllx=0pt,bblly=70pt,bburx=575pt,bbury=800pt,%
        width=15cm,angle=-90}
\vspace*{-0.5cm}

\vspace*{-1.1cm}
\hspace*{0.0cm}
\epsfig{file=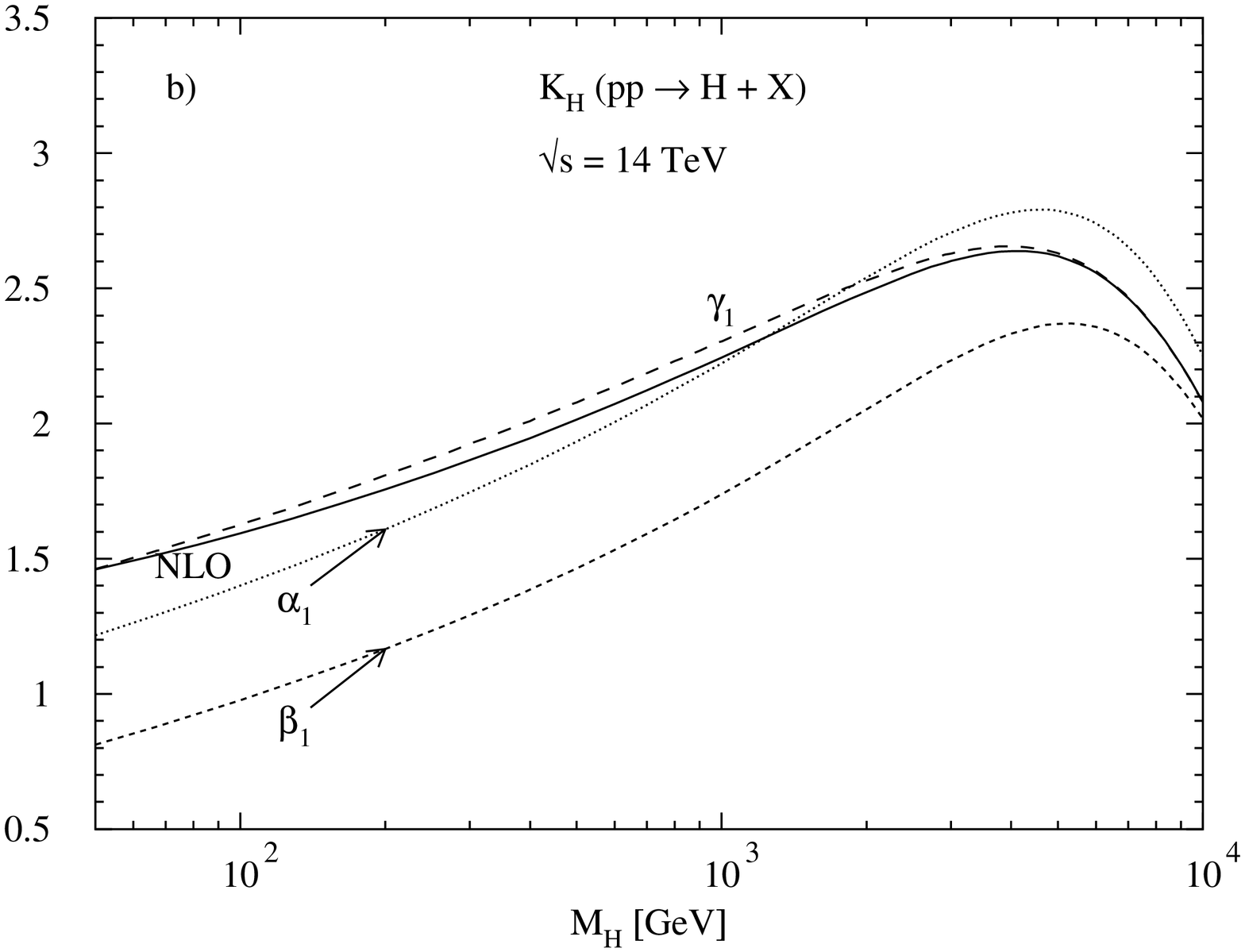,bbllx=0pt,bblly=70pt,bburx=575pt,bbury=800pt,%
        width=15cm,angle=-90}
\vspace*{-1.2cm}

\caption[]{\it a) Exact and approximate two- and three-loop partonic
  K-factors, convoluted with the NLO gluon-gluon luminosity $d{\cal
    L}^{gg}_{NLO}/d\tau$, in the heavy top-mass limit. The results for
  the three different schemes are presented as a function of the
  scalar Higgs mass $M_H$, using NLO CTEQ4M parton densities
  \cite{CTEQ4} and $\alpha_s$ [$\Lambda_{\msb}^{(5)} = 202$ MeV].  b)
  Hadronic NLO K-factor using LO CTEQ4L parton densities \cite{CTEQ4}
  and $\alpha_s$ [$\Lambda_{\rm LO}^{(5)} = 181$ MeV] for the LO cross
  section and including the NLO contributions from $\kappa_H$.}
\label{Kf1}
\end{figure}

\begin{figure}[htbp]
\vspace*{-1.1cm}
\hspace*{0.0cm}
\epsfig{file=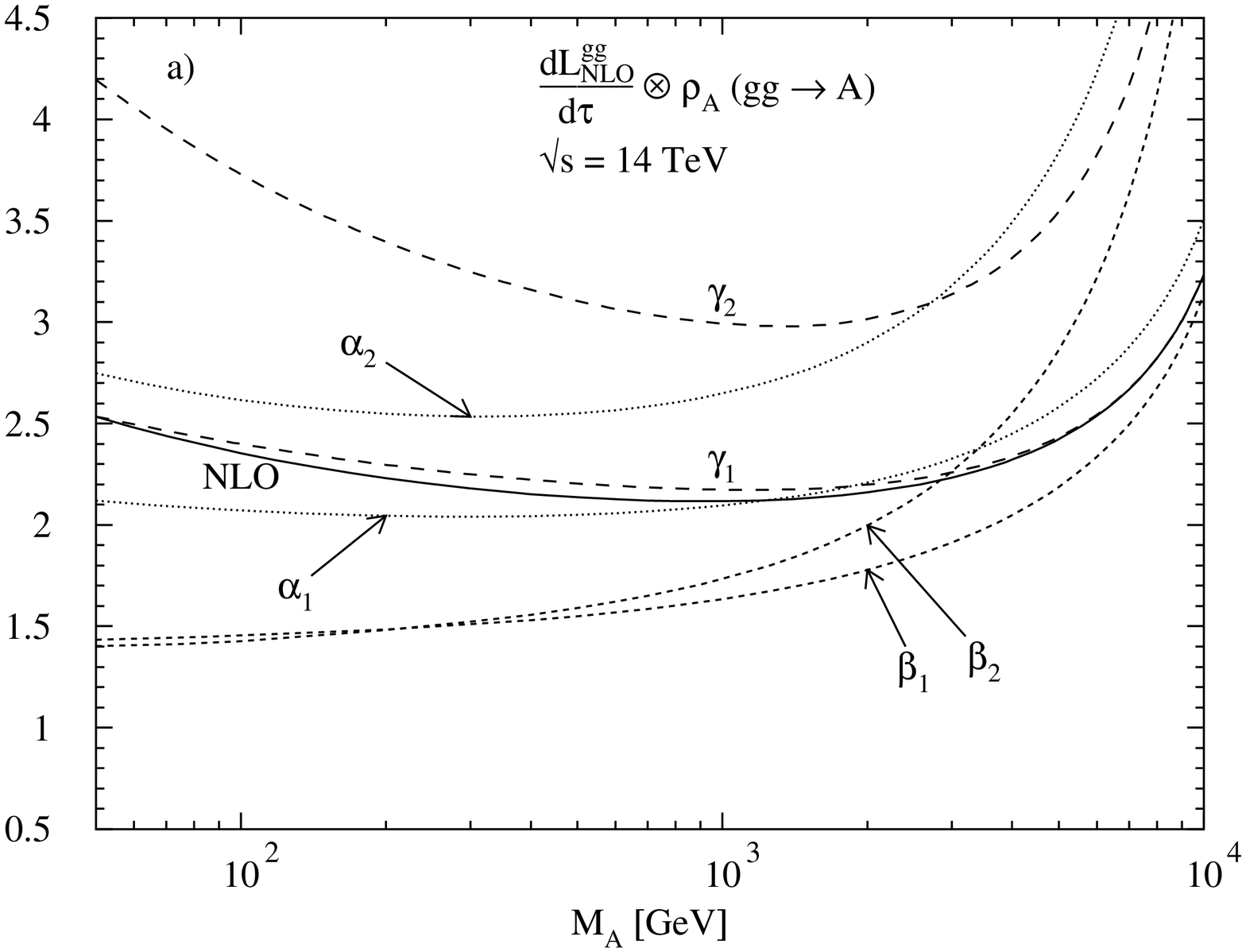,bbllx=0pt,bblly=70pt,bburx=575pt,bbury=800pt,%
        width=15cm,angle=-90}
\vspace*{-0.5cm}

\vspace*{-1.1cm}
\hspace*{0.0cm}
\epsfig{file=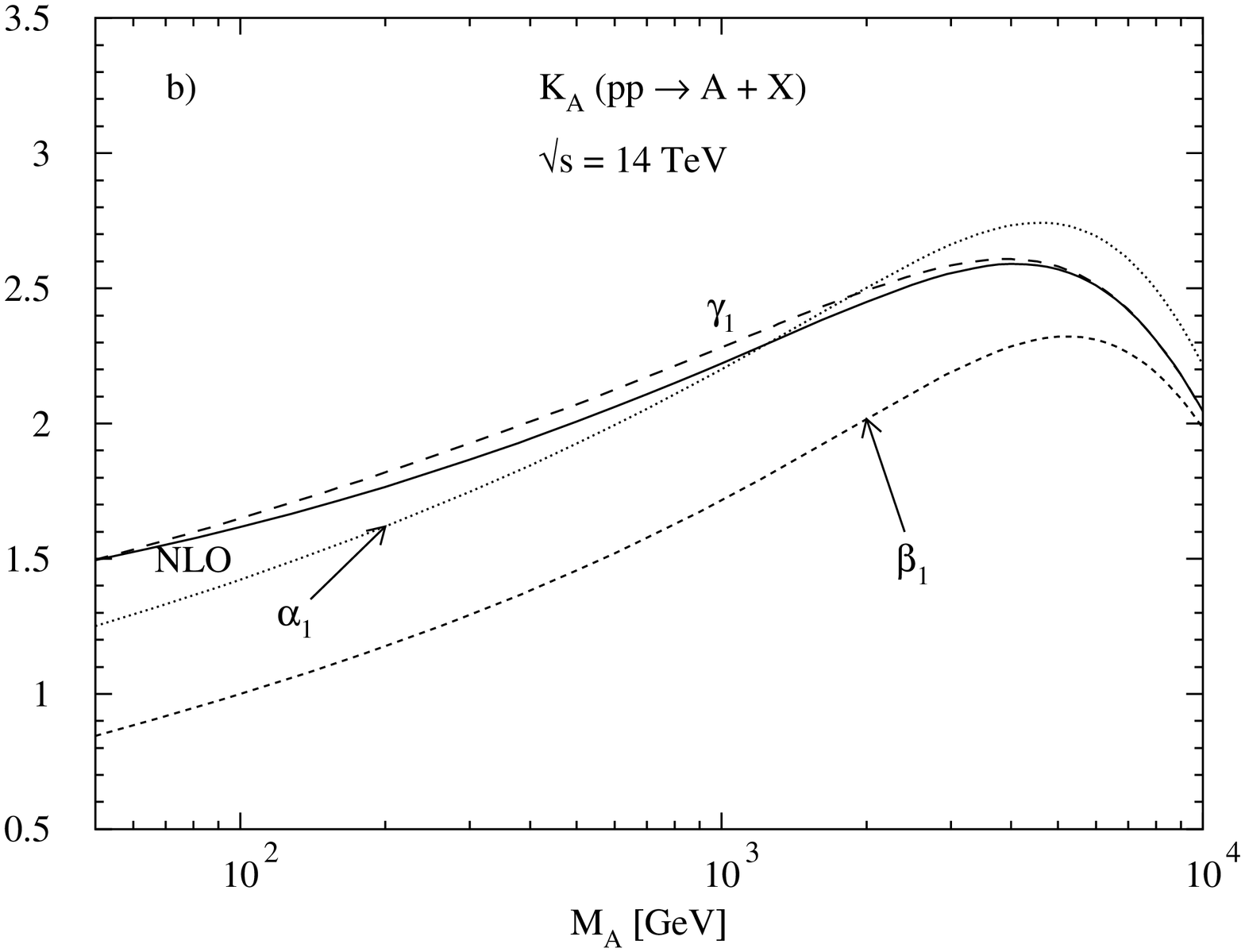,bbllx=0pt,bblly=70pt,bburx=575pt,bbury=800pt,%
        width=15cm,angle=-90}
\vspace*{-1.2cm}

\caption{\it As in the previous figure, but now for the pseudoscalar 
  Higgs in the MSSM for small \tb.}
\label{Kf2}
\end{figure}

Analogous to eq.~(\ref{sumpth}) the renormalized correction factor
$\rho_\phi$ of eqs.~(\ref{rhoa}-\ref{rhoA}) may be expanded as
\beq
\rho_\phi(z,Q^2/M^2,\alpha) = \sum_{n=0}^{\infty}~
\alpha^n(M^2) \rho_\phi^{(n)}(z,Q^2/M^2)
\label{rhoexp}
\eeq
We begin with the NLO expansion. Using the Mellin transform formulae
of the Appendix and the methods of \cite{Magnea}, we find that the
$O(\alpha)$ terms, upon inversion to momentum space [keeping the
renormalization/factorization scale $M^2$ different from $Q^2$] are
given by
\bea
\!\!\!\!\!\!
\rho_{h,H}^{\alpha (1)}(z,\frac{Q^2}{M^2})\!\!\!\! & = &\!\!\!\!
C_A\Big\{ 
4 {\cal D}_1(z) + 2 {\cal D}_0(z) L_M + (\frac{\pi^2}{3}+\frac{203}{36}
 - \frac{11}{6} L_M)\delta(1-z)
\Big\}, \label{rhoonea}\\
\!\!\!\!\!\!
\rho_{h,H}^{\beta (1)}(z,\frac{Q^2}{M^2})\!\!\!\! & = &\!\!\!\!
C_A\Big\{ 
4 {\cal D}_1(z) + 2 {\cal D}_0(z) L_M + \frac{\pi^2}{3}\delta(1-z)
\Big\}, \label{rhooneb}\\
\!\!\!\!\!\!
\rho_{h,H}^{\gamma (1)}(z,\frac{Q^2}{M^2})\!\!\!\! & = &\!\!\!\!
\rho_{h,H}^{\beta (1)}(z,\frac{Q^2}{M^2}) -
C_A\Big\{ 8{\cal E}_1(z) \Big\}\,, \label{rhoonec} \\
\!\!\!\!\!\!
\rho_A^{(1)}(z,\frac{Q^2}{M^2})\!\!\!\! & = &\!\!\!\!
\rho_{h,H}^{(1)}(z,\frac{Q^2}{M^2}) + 2 C_A \delta(1-z)
\hspace{2cm} \mbox{[for all three schemes]} \,,
\label{rhooneA}
\eea
where we use the notation
\beq
{\cal D}_i(z) = \left[\frac{\ln^i(1-z)}{1-z}\right]_+
\quad, \quad
{\cal E}_i(z) = \ln^i(1-z)
\quad, \quad
 L_M = \ln(Q^2/M^2)\,.
\eeq
With the results above we construct the correction factors according
to eq.~(\ref{rhoexp}) for the three schemes $\alpha$, $\beta$ and
$\gamma$ so that we can compare with exact results.  Similarly one may
obtain results to $O(\alpha^2)$. We find
\bea
\rho_{h,H}^{\alpha (2)}(z,Q^2/M^2) & = & 
C_A\Big\{
      8 C_A {\cal D}_3(z) +
      (-4 b_2 + 12 C_A L_M ) {\cal D}_2(z)  \nonumber\\ &+&
      (\frac{203}{9}C_A - 8 C_A \zeta_2 -4 b_2 L_M  -\frac{22}{3}C_A L_M
       + 4 C_A L_M^2) {\cal D}_1(z) 
           \nonumber\\ &+& 
      (16 C_A \zeta_3 -4C_A\zeta_2 L_M +\frac{203}{18}C_A L_M -
              b_2 L_M^2 -\frac{11}{3}C_A L_M^2){\cal D}_0(z)  \nonumber \\ &+& 
      (\frac{203}{18}C_A \zeta_2 + 6 C_A \zeta_2^2 -12C_A \zeta_4 + 
              \frac{41209}{2592}C_A  \nonumber\\&+& 
               8 C_A \zeta_3L_M -2\zeta_2 b_2 L_M-\frac{203}{36}b_2L_M
               -\frac{11}{3}C_A \zeta_2 L_M - \frac{2233}{216}C_A L_M
               \nonumber\\
               &+&\frac{121}{72}C_AL_M^2-2C_A\zeta_2 L_M^2
               +\frac{11}{12}b_2L_M^2) \delta(1-z) \Big\} \,,\label{rhotwoa}\\
\rho_{h,H}^{\beta (2)}(z,Q^2/M^2) & = & 
C_A\Big\{
      8 C_A {\cal D}_3(z) +
      (-4 b_2 + 12 C_A L_M ) {\cal D}_2(z)  \nonumber\\ &+&
      ( - 8 C_A \zeta_2  -4 b_2 L_M  + 
            4 C_A L_M^2) {\cal D}_1(z) \nonumber\\ &+& 
      (16 C_A \zeta_3 -4C_A\zeta_2 L_M -
              b_2 L_M^2){\cal D}_0(z)  \nonumber \\ &+& 
      (6 C_A \zeta_2^2 -12C_A \zeta_4 - \frac{2909}{216}b_2 
               -2\zeta_2 b_2 L_M \nonumber\\&+& 
            8 C_A \zeta_3L_M -
              2 C_A\zeta_2 L_M^2 ) \delta(1-z) \Big\} \,,
         \label{rhotwob}\\
\rho_{h,H}^{\gamma (2)}(z,Q^2/M^2) & = & 
\rho_{h,H}^{\beta (2)}(z,Q^2/M^2) + C_A\Big\{
       - 16 C_A {\cal E}_3(z)
              \nonumber \\ &+& 
       (8b_2 + 8C_A-24C_A L_M){\cal E}_2(z)\nonumber \\ &+&
        (16C_A \zeta_2+8b_2L_M+8C_AL_M-8C_A L_M^2){\cal E}_1(z) \Big\} \,,
\label{rhotwoc} \\
\rho_A^{\alpha (2)}(z,Q^2/M^2) \!\! & = & \!\!
\rho_{h,H}^{\alpha (2)}(z,Q^2/M^2)
+ C_A \Big\{ 8 C_A {\cal D}_1(z) + 4 C_A {\cal D}_0(z) L_M 
\nonumber\\
& + & (4 C_A \zeta_2 - \frac{11}{3} C_A L_M + \frac{239}{18} C_A 
- 2 b_2 L_M)\delta(1-z)\Big\} \,,
\label{rhotwoaA} \\
\rho_A^{\beta (2)}(z,Q^2/M^2) \!\! & = & \!\!
\rho_{h,H}^{\beta (2)}(z,Q^2/M^2)
+ C_A \Big\{ 8 C_A {\cal D}_1(z) + 4 C_A {\cal D}_0(z) L_M 
\nonumber\\
& + & (4 C_A \zeta_2 + 2 C_A - 2 b_2 L_M)\delta(1-z)\Big\} \,,
\label{rhotwobA} \\
\rho_A^{\gamma (2)}(z,Q^2/M^2) \!\! & = & \!\! 
\rho_{h,H}^{\gamma (2)}(z,Q^2/M^2)
+ C_A \Big\{ 8 C_A {\cal D}_1(z) + 4 C_A {\cal D}_0(z) L_M 
\nonumber\\
& - & 16 C_A {\cal E}_1(z)
+ (4 C_A \zeta_2 + 2 C_A - 2 b_2 L_M)\delta(1-z)\Big\} \,.
\label{rhotwocA}
\eea
In constructing the scale logarithms in the above expressions 
we have used the replacement
\begin{equation}
\alpha(Q^2) = \alpha(M^2) - \alpha(M^2)^2 b_2 L_M
\end{equation}
in the last exponents of eqs.~(\ref{rhoa},\ref{rhoA}).  We note that
this effectively amounts to including one term from the two-loop
Sudakov evolution kernel $W_\phi^{(2)}$.  This term can be derived
from the one-loop evolution kernel using the renormalization group,
see the discussion below eq.~(\ref{h1}).  In Fig.~\ref{Kf1} we present
the correction factors for SM Higgs production at the LHC, which
coincide with the correction factors of MSSM scalar Higgs boson
production for small \tb, and in Fig.~\ref{Kf2} for MSSM pseudoscalar
Higgs production, where we have identified $M^2=Q^2$.  For all results
in this section we used the CTEQ4M parton densities \cite{CTEQ4}, a
two-loop running coupling constant, with $\Lambda^{(5)}_{\msb} = 202$
MeV for the NLO and NNLO quantities and CTEQ4L densities with a LO
strong coupling constant ($\Lambda^{(5)}_{\rm LO} = 181$ MeV) for the
LO quantities.

In Figs.~\ref{Kf1}a and \ref{Kf2}a the ``partonic'' K-factors,
obtained from folding the correction factors $\rho_\phi$ with NLO
parton densities and using a NLO strong coupling for all orders of the
cross sections, are presented. For comparison we show in
Figs.~\ref{Kf1}b and \ref{Kf2}b the corresponding NLO ``hadronic''
K-factors [which include the small contributions from $\kappa_\phi$]
normalized to the LO cross sections evolved with LO parton densities
and $\alpha_s$.  Whereas the former indicate the effect of the higher 
order corrections to the partonic cross section, the latter exhibit
the convergence of the perturbative approach to the physical
(hadronic) quantities.  We observe from Figs.~\ref{Kf1} and \ref{Kf2}
that at NLO scheme $\gamma$ reproduces the exact NLO calculation
almost exactly for the full range of the SM Higgs mass $M_H \gsim 65$
GeV and the MSSM Higgs masses $M_\phi \gsim 45$ GeV, whereas the
schemes $\alpha$ and $\beta$ agree with the exact result only for
$M_\phi \gg 1$ TeV (the agreement of scheme $\alpha$ in the
intermediate Higgs mass range is accidental).  Moreover, note that the
NNLO corrections to the partonic cross sections in scheme $\gamma$ are
still significant.  Full NNLO predictions for hadronic cross sections
require NNLO parton densities, which are not yet available. At NLO, a
significant reduction of the hadronic K-factors compared to the
partonic K-factors can be read off from Figs.~\ref{Kf1}b and
\ref{Kf2}b, indicating a more reliable perturbative QCD expansion
contrary to what Figs.~\ref{Kf1}a and \ref{Kf2}a suggest.  We point
out that should the size of the NNLO corrections to the physical cross
section warrant concern about the convergence of the perturbative
approach, our resummation method can provide a tool to control such
large corrections.

Besides the three-loop anomalous dimension, the determination of NNLO
densities requires NNLO calculations for the physical quantities
included in a global fit. However, there are presently only a few
exact NNLO calculations available \cite{DYexact,DIS,ZN2}. An
approximate way to proceed might be provided by the use of NNLO
expansions of resummed cross sections, which at NLO have to
approximate the exact results reliably. For future high-energy hadron
colliders we expect this to require the inclusion of the novel
subleading contributions that have been discussed in our analysis.
Once approximate NNLO results have been obtained for several
processes, e.g.~heavy flavor production at the LHC, and the three-loop
anomalous dimensions for the NNLO evolution of the parton densities
have been calculated, a global fit of approximate NNLO parton
densities can be performed.  The same procedure could of course also
be followed at the resummed level.

\begin{figure}[htb]

\vspace*{-1.1cm}
\hspace*{0.0cm}
\epsfig{file=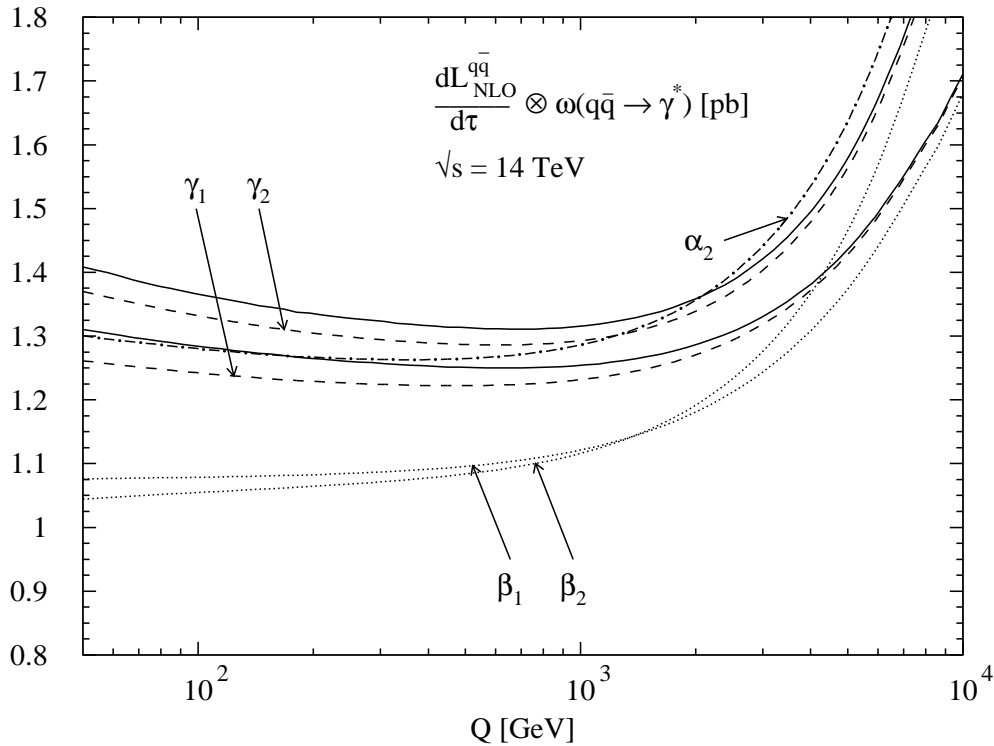,bbllx=0pt,bblly=70pt,bburx=575pt,bbury=800pt,%
        width=15cm,angle=-90}
\vspace*{-1.2cm}

\caption[]{\it Exact and approximate one- and two-loop
  partonic K-factors, using NLO CTEQ4M parton densities \cite{CTEQ4}
  and strong coupling [$\Lambda_{\msb}^{(5)} = 202$ MeV] in all orders
  of the cross section, for Drell-Yan, in three different schemes, as
  a function of the $\gamma^*$ mass $Q$. The lower solid line is the
  exact NLO result, in the $q\bar{q}$ channel, and the upper solid
  line is the NNLO one.}
\label{KfM}
\end{figure}

To investigate the reliability of scheme $\gamma$ at NNLO we confront
in Fig.~\ref{KfM} the approximate partonic K-factor, using NLO parton
densities and strong coupling in all expressions, for the Drell-Yan
$q\bar{q}$ production channel at the LHC as a function of the
off-shell photon mass $Q$ with the exact calculation of
Ref.~\cite{DYexact}.  We do this at NLO and NNLO in schemes $\beta$
and $\gamma$. The expressions used are given in eqs.~(\ref{hpdyb}) and
(\ref{hpdyg}) of the Appendix.  We observe that for scheme $\gamma$
the agreement is again excellent at NLO. As for the Higgs case, this
is perhaps not so surprising, since the NLO answer is to a large
extent included in the evolution kernel. However, note that the
agreement is remarkably good even at NNLO. Clearly the $\ln^i(1-z)$
terms are dominant in this order as well.  In scheme $\alpha$ we found
at NLO a similar accidental agreement with the exact result, as in the
Higgs case. In Fig.~\ref{KfM} we show the NNLO result for scheme
$\alpha$, which fails to approximate the exact NNLO curve, in contrast
with scheme $\gamma$.  Note that scheme $\beta$ fails at medium and
low $Q^2$.  The impressive agreement in NNLO, in combination with the
general arguments given at the end of the previous section, gives us
confidence to assert that the NNLO expansion for Higgs production in
scheme $\gamma$ will be close to the exact value.

\begin{figure}[htbp]

\vspace*{-1.5cm}
\hspace*{-0.1cm}
\epsfig{file=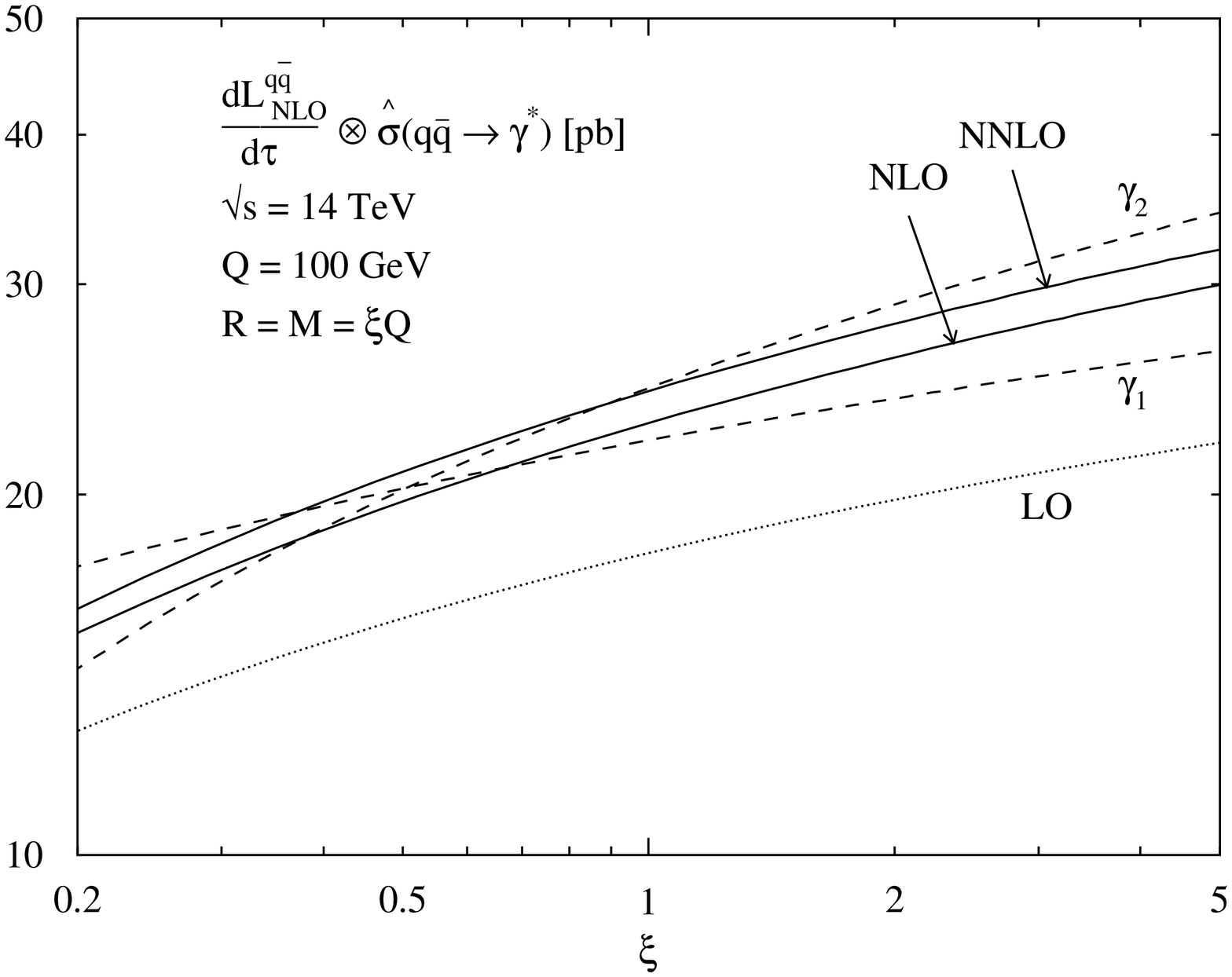,bbllx=0pt,bblly=70pt,bburx=575pt,bbury=800pt,%
        width=15cm,angle=-90}
\vspace*{-1.1cm}

\epsfig{file=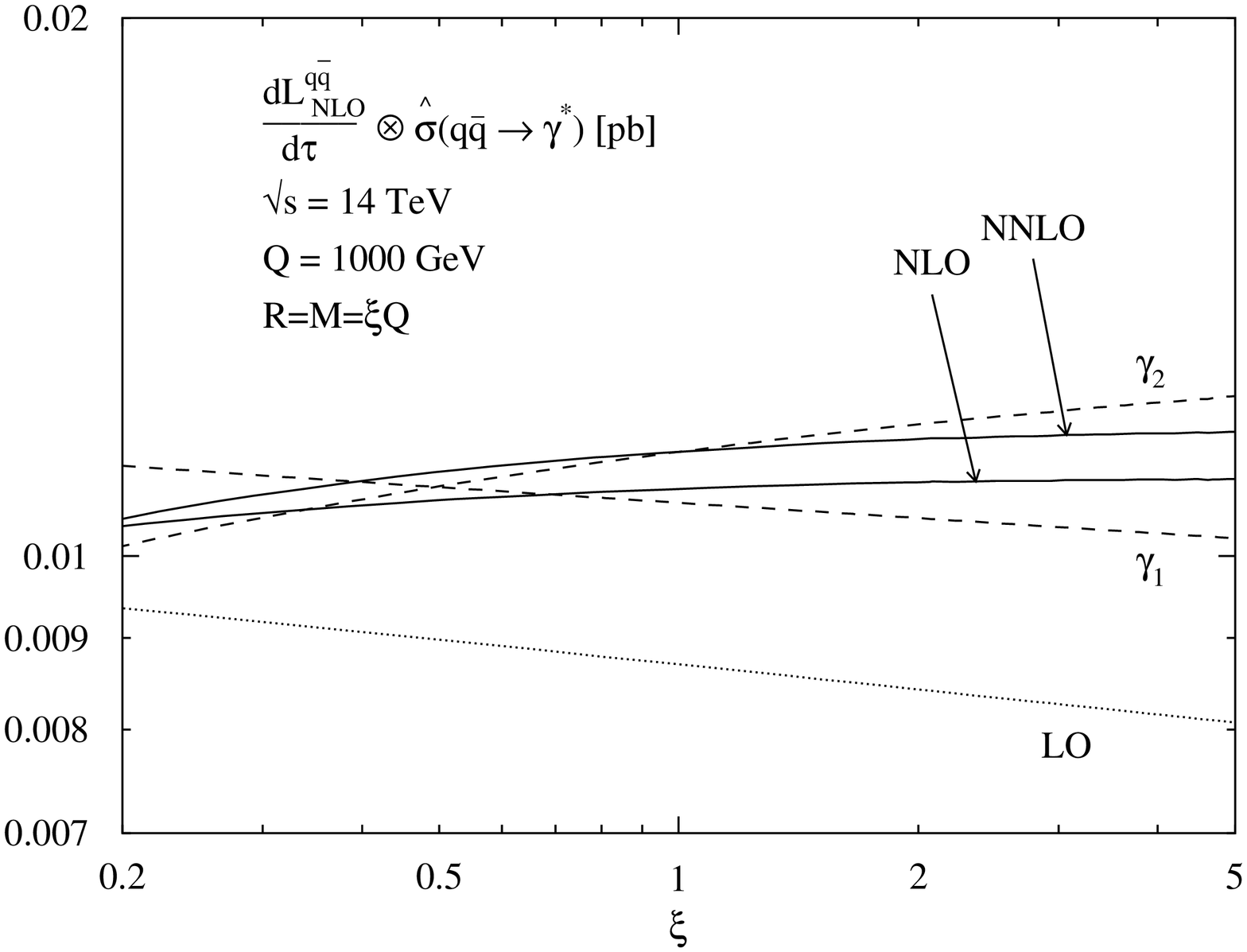,bbllx=0pt,bblly=70pt,bburx=575pt,bbury=800pt,%
        width=15cm,angle=-90}
\vspace*{-1.0cm}

\caption[]{\it Scale dependence of the Drell-Yan cross section for two
  values of the $\gamma^*$ mass $Q$. The solid lines represent the
  exact calculation at NLO and NNLO and the dotted line the LO one.
  NLO CTEQ4M parton densities \cite{CTEQ4} and strong coupling
  [$\Lambda_{\msb}^{(5)} = 202$ MeV] have been used.}
\label{fg:dy_sc}
\end{figure}

Next we investigate the consequences of the scale logarithms for Higgs
production. It was found in Ref.\cite{SDGZ} that the scale dependence
of the NLO cross section is still a monotonous function of the scales.
In view of the outstanding agreement of the exact results with our
approximate ones in scheme $\gamma$ for $M^2 = Q^2$ for the
$\tau=Q^2/S$ dependence, we may use the same results to examine the
scale dependence at NNLO.  In fact, from arguments such as given in
\cite{DYRG}, one may deduce that eqs.~(\ref{rhotwoa}-\ref{rhotwocA})
approximate the exact scale dependent terms very well, the only term
lacking being proportional to the two-loop anomalous dimension
eq.~(\ref{W2}), which we have omitted.  Again we use the Drell-Yan
case to gauge the quality of scheme $\gamma$ in describing the scale
dependence. This is done in Fig.~\ref{fg:dy_sc} for two values of the
$\gamma^*$ mass $Q$.  To obtain the best approximation, we have
defined the curve labelled $\gamma_2$, the NNLO approximation in
scheme $\gamma$, by adding the NNLO term of the expanded resummed
exponential to the exact NLO result, rather than the $\gamma_1$ curve.
We see that at NLO and NNLO scheme $\gamma$ again describes the scale
dependence quite reasonably.  Again, for both the exact and
approximate expressions a consistent analysis requires the use of NNLO
parton densities in determining the scale dependence, but these are
not available yet.  (It was shown in \cite{DYexact} that for the full
NNLO Drell-Yan cross section, including other production channels,
there is a strong indication, albeit based on NLO parton densities,
that the scale dependence is significantly reduced compared to NLO).

\begin{figure}[htbp]

\vspace*{-1.5cm}
\hspace*{-0.2cm}
\epsfig{file=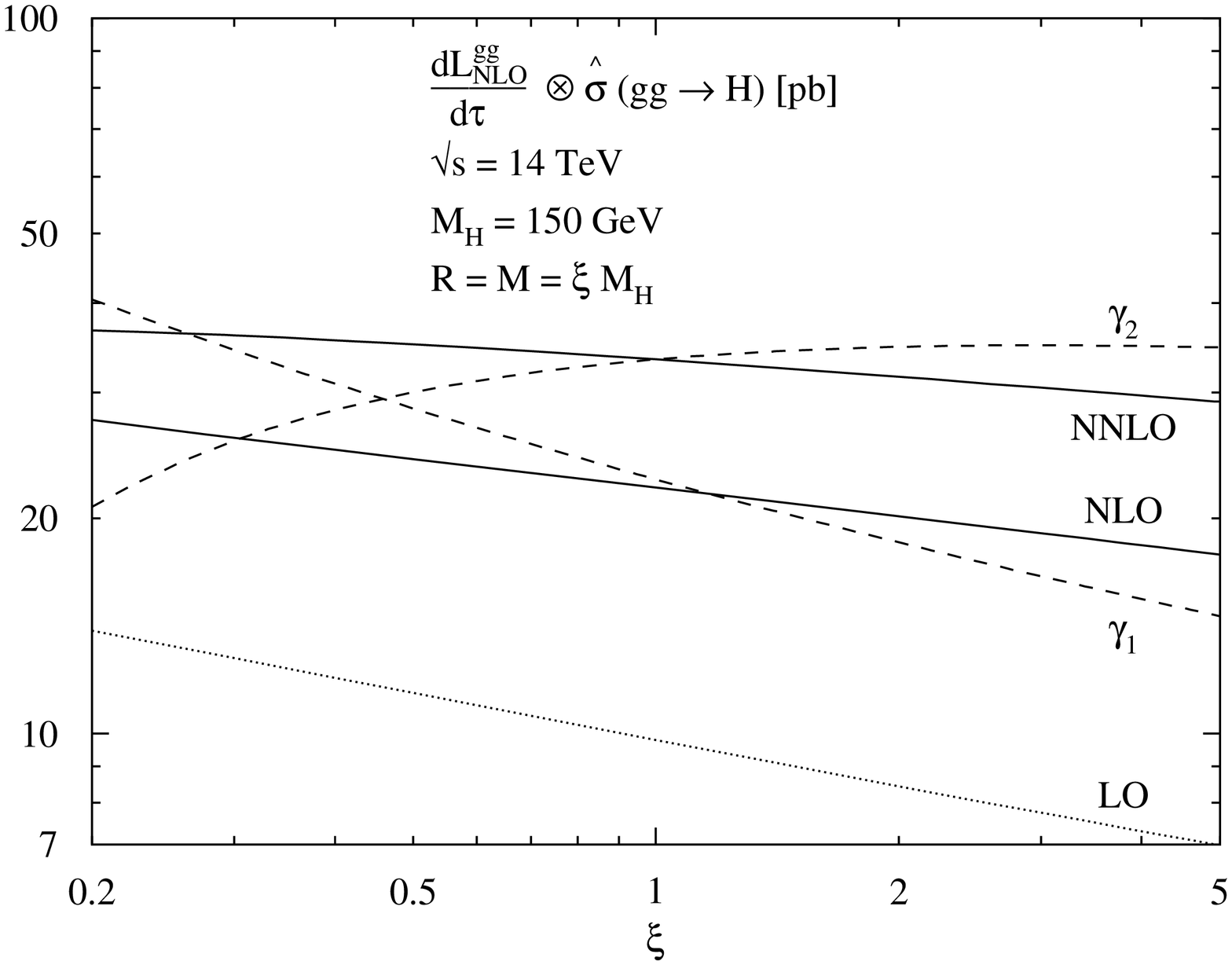,bbllx=0pt,bblly=70pt,bburx=575pt,bbury=800pt,%
        width=15cm,angle=-90}
\vspace*{-1.1cm}

\epsfig{file=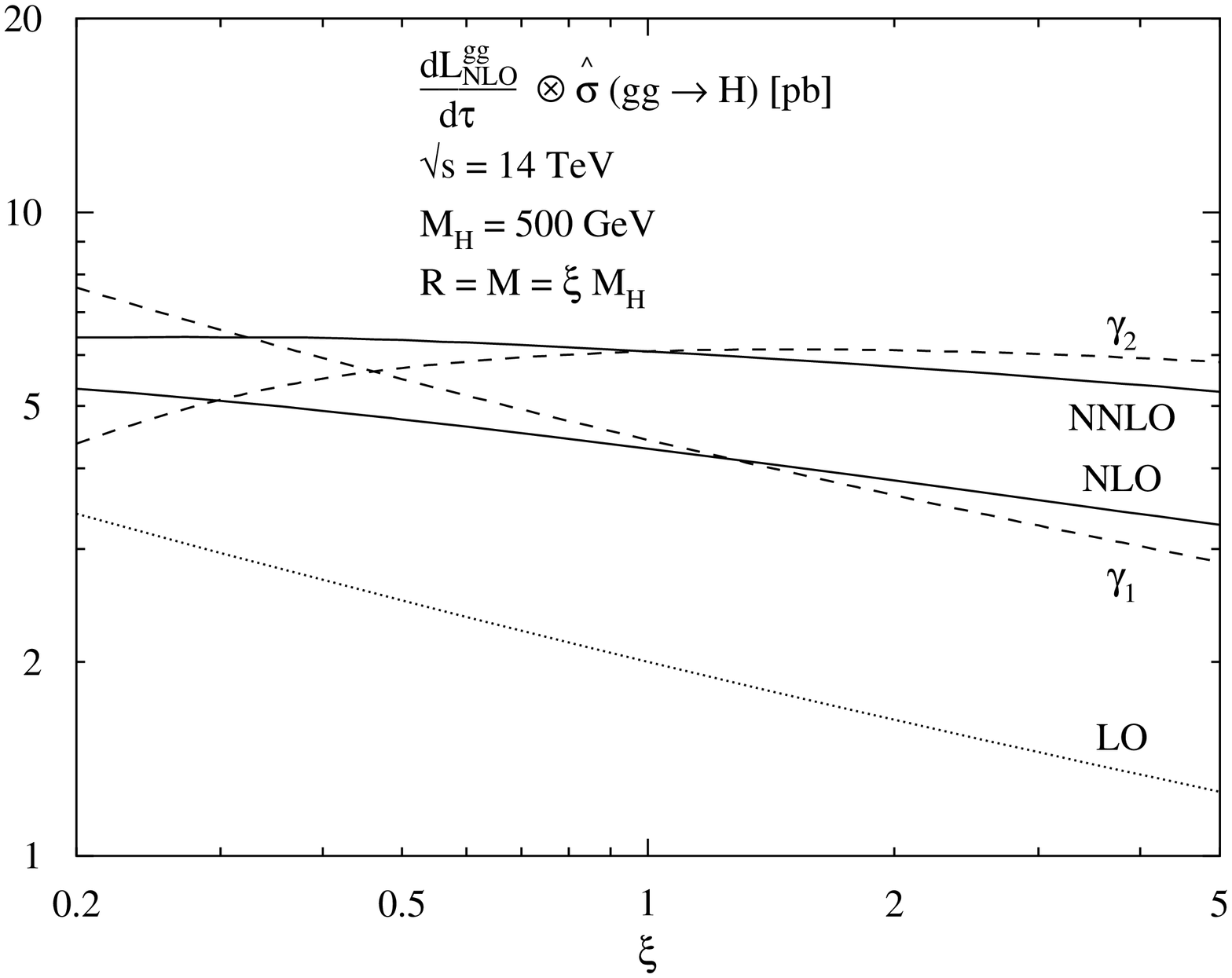,bbllx=0pt,bblly=70pt,bburx=575pt,bbury=800pt,%
        width=15cm,angle=-90}
\vspace*{-1.0cm}

\caption[]{\it Scale dependence of the Higgs production cross section for two
  values of the Higgs mass $M_H$.  NLO CTEQ4M parton densities
  \cite{CTEQ4} and strong coupling [$\Lambda_{\msb}^{(5)} = 202$ MeV]
  have been used in all expressions, so that the NNLO results do not
  correspond to the physical NNLO cross sections.}
\label{fg:Higgs_sc}
\end{figure}
In Fig.~\ref{fg:Higgs_sc} we present the scale dependence for SM Higgs
production.  The curve labelled $\gamma_1$ includes the sum of the
approximate term $\rho_H^{\gamma (1)}$ of eq.~(\ref{rhoonec}) and the
NLO contribution to $\kappa_H$ of eq.~(\ref{kappa}) for the $gg$
initial state.  The approximate NNLO result, labelled $\gamma_2$, has
been obtained by adding the NNLO term $\rho_H^{\gamma (2)}$ of
eq.~(\ref{rhotwoc}) and the corresponding contributions up to NNLO to
$\kappa_H$ of eq.~(\ref{kappa}) to the exact NLO result [this
significantly improves the approximation similar to the Drell-Yan
process].  The curves labelled NNLO include the full NNLO scale
dependence in the partonic cross section, which has been obtained from
the exact NLO result by means of renormalization group methods,
neglecting quarks at all stages. This curve has been normalized to the
$\gamma_2$ curve at $\xi=1$.  We observe that the NLO term of scheme
$\gamma$ deviates from the exact NLO slightly at large scales and
significantly for small scales. This is caused by terms of ${\cal
  O}(1/N)$, which have been neglected.  At NNLO there is a strong
indication for a significant stabilization of the theoretical
prediction for the total Higgs production cross section at the LHC.
There are deviations between the NNLO and $\gamma_2$ curves
at small and large scales, which are again due to terms of ${\cal
  O}(1/N)$ that have been neglected in scheme $\gamma$.
We found similar results to hold for the MSSM pseudoscalar Higgs case.

Finally, let us comment on the phenomenological implications of our
results for the Higgs K-factor at NNLO. When the NLO corrections to
the Higgs production cross section were calculated both for the
infinite mass limit \cite{S:limit} and for the general massive case
\cite{SDGZ} it was found that the ratio of the NLO cross section to
the LO one could be larger than two if one used NLO parton densities
and strong coupling for both cross sections at LO and NLO.  In order
to estimate the increase in size of the QCD corrected physical cross
section the hadronic K-factor has to be defined by including the
corresponding cross sections evaluated with parton densities and
strong coupling at the {\it same} order, i.e.~LO cross sections with
LO quantities and NLO with NLO quantities.  This hadronic NLO K-factor
amounts to 1.5-2.0 in the phenomenologically relevant Higgs mass
range.  This indicates that the procedure to predict with increasing
accuracy the physical cross section, by including higher order
corrections consistently in all quantities entering the factorization
theorem, seems to converge better than one might conclude from
Figs.~\ref{Kf1}a and \ref{Kf2}a.  A consistent NNLO hadronic K-factor
requires NNLO strong coupling and parton densities, which are not yet
available, but one might expect a further reduction from them.
However, as long as we do not know the parton densities beyond NLO, we
observe from Figs.~\ref{Kf1}a and \ref{Kf2}a that NNLO corrections to
the partonic cross sections are sizable.  Our demonstration that the
bulk of the corrections originates from soft and collinear gluon
radiation, and the fact that they can be resummed analytically,
provides then a different way to organize the perturbative expansion
in the phenomenologically relevant Higgs mass range: one may redefine
the original QCD perturbation series by rewriting it as the product of
our resummed expression, times a new series. Due to the excellent
approximation of the original series by the expanded resummed series,
the new series is expected to be very well behaved perturbatively.
This is of course the standard method for making sense of perturbative
QCD near the edges of phase space, where QCD corrections are large.
However, the result can now be extended much further away from
threshold due to the inclusion of the novel subleading contributions.
Note that this procedure would also require resummed parton densities.
We furthermore note that the large size of the corrections compared to
Drell-Yan is partly due to the fact that for every power of $\alpha_s$
a color factor $C_A=3$ appears. We have seen that for Drell-Yan in the
$\msb$ scheme, where the corresponding color factor is $C_F=4/3$ but
the analytical structure of the soft gluon corrections is quite
similar, the corrections are considerably smaller. The same phenomenon
was observed for the case of resummed heavy quark production in
Ref.~\cite{LSN}. These color factors are correctly included in the
resummed formulae.
 
As remarked earlier, the evaluation of the resummed series has its own
subtleties, related to the appearance of the infrared renormalon. For
its treatment there have recently been a number of proposals
\cite{renormalon}, which we will not discuss here.  We anticipate that
the resummed Higgs production cross section will be quite sensitive to
the details of handling the renormalon, due the large color factor in
the exponent \cite{LSN}.

\section{Conclusions}

In this paper we have performed the all order resummation of soft
gluon effects in Higgs production both for the Standard Model and its
minimal supersymmetric extension, to next-to-leading logarithmic
accuracy.  We have extended the usual resummation formalism to include
logarithms which, although integrable, diverge in the partonic cross
section near the edge of phase space.  By expanding our resummed
results to NLO and NNLO, and using the Drell-Yan process for
comparison, we have shown that this extension expands the
applicability of resummation efforts into the phenomenologically
relevant Higgs boson production range at the LHC.  An accurate
assessment of the expected Higgs production rate is of paramount
importance for the LHC physics programme. However, a physical
prediction of the NNLO cross section requires knowledge of NNLO parton
densities, which is not yet available.  Clearly, in this regard, it
would be interesting to investigate the applicability of our extended
formalism to many other QCD production processes with potentially
large K-factors at NLO, e.g. heavy quark production \cite{HQ} both at
the Tevatron and the LHC, or to revisit the Drell-Yan process for
phenomenological studies along the lines of \cite{DYphen}.

\subsection*{Acknowledgements}

We would like to thank K.~Chetyrkin, S.A.~Larin, P.~Nason, J.~Smith,
G.~Sterman and P.M.~Zerwas for helpful conversations.  We also thank 
W.~van Neerven and P.~Rijken for providing the fortran code of the 
NNLO Drell-Yan process. E.L. would like to thank the Columbia University 
theory group and M.K. the CERN Theory Division for hospitality while 
this work was being completed.

\appendix

\section{Useful formulae}

In this appendix we collect some useful formulae used in section
2. We begin by extending the Mellin transform table of Ref.~\cite{CT}
up to $O(1/N)$.
Define
\beq
I_n(N) = \int_0^1~dx~ x^{N-1}\ \Big[\frac{\ln^n(1-x)}{1-x}\Big]_+.
\eeq
For the lowest four values of $n$ this integral is, up
to $O(1/N)$
\bea
I_0(N) &=& -\lnNt + \frac{1}{2}\frac{1}{N} \label{i0}\\
I_1(N) &=& \frac{1}{2}\ln^2\Nt + \frac{1}{2}\zeta_2 
   -\Big(\frac{1}{2}\lnNt+\frac{1}{2}\Big)\frac{1}{N} \\
I_2(N) &=& -\frac{1}{3}\ln^3\Nt -\zeta_2\lnNt -\frac{2}{3}\zeta_3
  +\Big(\frac{1}{2}\ln^2\Nt + \frac{1}{2}\zeta_2+\lnNt\Big)\frac{1}{N}\\
I_3(N) &=& \frac{1}{4}\ln^4\Nt + \frac{3}{2}\zeta_2\ln^2\Nt
 + 2\zeta_3\lnNt +\frac{3}{4}\zeta_2^2+\frac{3}{2}\zeta_4 \nonumber\\
&+& \Big(-\frac{1}{2}\ln^3\Nt-\frac{3}{2}\zeta_2\lnNt
   -\zeta_3-\frac{3}{2}\ln^2\Nt
  -\frac{3}{2}\zeta_2\Big)\frac{1}{N}
\eea
with $\Nt = N e^{\gamma_E}$ and $\gamma_E$ denoting the Euler constant.
Define also
\beq
J_n(N) = \int_0^1~dx~x^{N-1}\ \ln^n(1-x)
\eeq
For the lowest four values of $n$ this integral is, up
to $O(1/N)$
\bea
J_0(N) &=& \frac{1}{N} \\
J_1(N) &=& -\frac{\ln \Nt}{N}\\
J_2(N) &=& \frac{\ln^2 \Nt}{N} + \frac{\zeta_2}{N}\\
J_3(N) &=& -\frac{\ln^3\Nt}{N} -3\zeta_2 \frac{\ln\Nt}{N}-2\frac{\zeta_3}{N}
\label{j3}
\eea
Next we present the NNLO perturbative expansions of the resummed hard
part $\omega_{q\bar{q}}$ of the $\msb$ Drell-Yan cross section in two
schemes, defined in analogy to eq.~(\ref{schemes}).  The relevant
function to approximate here is $(z^{N-1}-1)g_{\rm
  DY}^{(1)}(z,\epsilon)$ with $g_{\rm DY}^{(1)}(z,\epsilon) =
C_F\,(1+z^2)$.  The schemes $\alpha$, $\beta$ and $\gamma$ are defined
by the replacements
\bea
\mbox{scheme~} \alpha: \frac{1}{C_F}
(z^{N-1}-1)g_{\rm DY}^{(1)}(z,\ep) & \go &
(z^{N-1}-1)~2  \nonumber \\
\mbox{scheme~} \beta: \frac{1}{C_F}
(z^{N-1}-1)g_{\rm DY}^{(1)}(z,\ep) & \go &
(z^{N-1}-1)~2 + (1-z)(1+z) \nonumber\\
\mbox{scheme~} \gamma: \frac{1}{C_F}
(z^{N-1}-1)g_{\rm DY}^{(1)}(z,\ep) & \go & 
(z^{N-1}-1)~2 + (1-z)(1+z) \nonumber \\
&& \hspace{2.35cm} - 2 z^{N-1}(1-z)\,. 
\eea
We find the results
\bea
\omega^{\msb}_{\alpha,q\bar{q}}(z,Q^2/M^2) & = & \delta(1-z) + 
\alpha(M^2)C_F\Big\{4{\cal D}_1(z) + 2 {\cal D}_0(z)L_M \nonumber\\ &+&
      (2\zeta_2-\frac{1}{2})\delta(1-z)\Big\}+\alpha(M^2)^2C_F\Big\{
      8 C_F {\cal D}_3(z) +
      (-4 b_2 \nonumber\\ &+& 12 C_F L_M ) {\cal D}_2(z)  +
      ( - 8 C_F \zeta_2 -2 C_F  -4b_2L_M + 
            4 C_F L_M^2) {\cal D}_1(z) \nonumber\\ &+& 
      (16 C_F \zeta_3 -4C_F\zeta_2 L_M  -b_2 L_M^2 - C_F L_M
              ){\cal D}_0(z)  \nonumber \\ &+& 
      (6 C_F \zeta_2^2 -12C_F \zeta_4 -C_F\zeta_2 +C_F\frac{1}{8}
               + \frac{1}{2}b_2 L_M -2\zeta_2 b_2 L_M 
             \nonumber\\&+& 8 C_F \zeta_3L_M -  
              2 C_F\zeta_2 L_M^2 ) \delta(1-z) \Big\} 
\label{hpdya} \\
\omega^{\msb}_{\beta,q\bar{q}}(z,Q^2/M^2) & = & \delta(1-z) + 
\alpha(M^2)C_F\Big\{4{\cal D}_1(z) + 2 {\cal D}_0(z)L_M \nonumber\\ &+&
      (2\zeta_2-4+\frac{3}{2}L_M)\delta(1-z)\Big\} \nonumber\\ &+&
\alpha(M^2)^2C_F\Big\{
      8 C_F {\cal D}_3(z) +
      (-4 b_2 + 12 C_F L_M ) {\cal D}_2(z)  \nonumber\\ &+&
      ( - 8 C_F \zeta_2 -16 C_F + 6 C_F L_M  -4b_2L_M + 
            4 C_F L_M^2) {\cal D}_1(z) \nonumber\\ &+& 
      (16 C_F \zeta_3 -4C_F\zeta_2 L_M  -8 C_FL_M -b_2 L_M^2
             + 3C_F L_M^2 ){\cal D}_0(z)  \nonumber \\ &+& 
      (6 C_F \zeta_2^2 -12C_F \zeta_4 - \frac{15}{2}b_2 -8C_F\zeta_2 + 8 C_F
               -6 C_F L_M\nonumber\\&+& 4b_2 L_M -2\zeta_2 b_2 L_M +
            3 C_F \zeta_2 L_M +8 C_F \zeta_3L_M 
               \nonumber\\&-&
              \frac{3}{4}b_2 L_M^2 -2 C_F\zeta_2 L_M^2 
              +\frac{9}{8}C_F L_M^2) \delta(1-z) \Big\} 
\label{hpdyb} \\
\omega^{\msb}_{\gamma,q\bar{q}}(z,Q^2/M^2) & = & 
\omega_{\beta,q\bar{q}}(z,Q^2/M^2) +
\nonumber\\ &+&
\alpha(M^2)C_F\Big\{ -4 {\cal E}_1(z) \Big\} \nonumber\\ &+&
\alpha(M^2)^2C_F\Big\{-8C_F {\cal E}_3(z) + 
 (4b_2 + 8C_F-12C_F L_M){\cal E}_2(z)\nonumber \\ &+&
        (16C_F +8 C_F\zeta_2 + 2C_F L_M +4b_2L_M-4C_F L_M^2){\cal E}_1(z)
         \Big\}.
\label{hpdyg}
\eea


\end{document}